\documentclass[%
 aip,
 amsmath,amssymb,
 reprint,%
]{revtex4-1}

\usepackage{graphicx}
\usepackage{dcolumn}
\usepackage{bm}
\usepackage[acronym]{glossaries}
\usepackage[utf8]{inputenc}
\usepackage[T1]{fontenc}
\usepackage{mathptmx}
\usepackage{etoolbox}
\usepackage{hyperref}

\usepackage{xcolor}

\newacronym{dft}{DFT}{density functional theory}
\newacronym{lrtddft}{LR-TDDFT}{linear-response time-dependent density functional theory}
\newacronym{paw}{PAW}{projector augmented wave}
\newacronym{tdm}{TDM}{transition dipole moment}
\newacronym{dm}{DM}{dipole moment}
\newacronym{1rtdm}{1-RTDM}{one-particle reduced transition density matrix}
\newacronym{1rdm}{1-RDM}{one-particle reduced density matrix}
\newacronym{ao}{AO}{atomic orbital}
\newacronym{mo}{MO}{molecular orbital}
\newacronym{lcao}{LCAO}{linear combination of atomic orbitals}
\newacronym{fd}{FD}{finite difference}
\newacronym{pw}{PW}{plane wave}
\newacronym{xc}{xc}{exchange-correlation}
\newacronym{ks}{KS}{Kohn-Sham}
\newacronym{oo}{OO}{orbital optimized}
\newacronym{mom}{MOM}{maximum overlap method}
\newacronym{homo}{HOMO}{highest occupied molecular orbital}
\newacronym{lumo}{LUMO}{lowest unoccupied molecular orbital}
\newacronym{si}{SI}{supporting information}
\newacronym{cbs}{cbs}{complete basis set}
\newacronym{roks}{ROKS}{restricted open-shell Kohn-Sham}
\newacronym{sd}{SD}{Slater Determinant}
\newacronym{tda}{TDA}{Tamm-Dancoff approximation}
\newacronym{cebe}{CEBE}{core electron binding energies}
\newacronym{hf}{HF}{Hartree-Fock}
\newacronym{sgm}{SGM}{square-gradient minimization}
\newacronym{xps}{XPS}{X-ray photoelectron spectroscopy}
\newacronym{xas}{XAS}{X-ray absorption spectroscopy}
\newacronym{fc}{FC}{Franck-Condon}
\newacronym{csf}{CSF}{configuration state function}
\newacronym{rmse}{RMSE}{root-mean-square error}
\newacronym{mse}{MSE}{mean signed error}
\newacronym{ci}{CI}{configuration interaction}

\newcommand{\ket}[1]{\vert #1 \rangle}
\newcommand{\bk}[2]{\langle #1 \vert #2 \rangle}

\newcommand{\expv}[3]{\langle #1 \vert #2 \vert #3 \rangle}

\newcommand{\mr}[1]{\mathrm{#1}}

\newcommand*{\citen}{}
\DeclareRobustCommand*{\citen}[1]{%
  \begingroup
    \romannumeral-`\x 
    \setcitestyle{numbers}%
    \cite{#1}%
  \endgroup
}

\usepackage[authormarkuptext=name,authormarkup=none]{changes}
\definecolor{red}{RGB}{255, 0, 0}
\definecolor{blue}{RGB}{0, 0, 255}
\definecolor{green}{RGB}{0, 192, 0}
\definechangesauthor[name={Authors}, color={blue}]{GL}

\makeatletter
\def\@email#1#2{%
 \endgroup
 \patchcmd{\titleblock@produce}
  {\frontmatter@RRAPformat}
  {\frontmatter@RRAPformat{\produce@RRAP{*#1\href{mailto:#2}{#2}}}\frontmatter@RRAPformat}
  {}{}
}%
\makeatother
\begin{document}

\preprint{AIP/123-QED}

\title{Orbital-optimized density functional calculations of excited electronic states: Recent advances and perspectives}
\author{Lorenzo Restaino}
 \affiliation{Science Institute and Faculty of Physical Sciences, University of Iceland, 107 Reykjav\'ik, Iceland}
\author{Giulia Gamboni}
 \affiliation{Department of Chemical and Pharmaceutical Sciences, University of Trieste, 34127 Trieste, Italy}
\author{Elli Selenius}
 \affiliation{Science Institute and Faculty of Physical Sciences, University of Iceland, 107 Reykjav\'ik, Iceland}
\author{Gianluca Levi}
 \affiliation{Department of Chemical and Pharmaceutical Sciences, University of Trieste, 34127 Trieste, Italy}
 \affiliation{Science Institute and Faculty of Physical Sciences, University of Iceland, 107 Reykjav\'ik, Iceland}
 \email{gianluca.levi@units.it}

\date{\today}

\begin{abstract}
Orbital-optimized (OO) density functional calculations provide a time-independent, variational route to electronic excitations, alternative to presently widely used time-dependent density functional theory (TDDFT) approaches. As the orbitals are optimized in a state specific way, these methods can provide a balanced description of excited states with different character, thereby overcoming several limitations of practical implementations of TDDFT. Driven by recent developments in algorithms for obtaining excited states as saddle points on the electronic energy surface, OO methods have attracted increasing interest, maturing into an active and rapidly expanding area of research. Here, the theoretical foundations of the approach are clarified and an overview of recent methodological developments in excited-state orbital optimization is provided. An overview of methods for treating open-shell singlet excited states and current approaches for computing transition properties and spectra is also provided. Finally, recent applications to molecular Rydberg, charge-transfer, and core excitations are reviewed, with the aim of assessing the present accuracy and range of applicability of OO density functional calculations with common exchange and correlation functionals.
\end{abstract}

\maketitle

\tableofcontents

\section{Introduction}
\Gls{dft}\cite{Kohn1965} has become one of the most widely used electronic structure methods for calculations of the ground electronic state\cite{Teale2022}, mainly thanks to the exceptionally favorable balance between accuracy and computational cost of its \gls{ks} formulation\cite{Kohn1965}. Decades of development of density functional approximations and efficient algorithms have made \gls{ks} \gls{dft} a routine tool for computing the structure and dynamics of molecular and condensed-phase systems in the ground state and enabled large-scale simulations that are for the moment beyond the reach of higher-level wave function methods\cite{Tanaka2024, Lin2021, Ko2020}.

Achieving a comparable level of applicability for excited electronic states has proven significantly difficult. Time-dependent DFT (TDDFT) is a formally exact theory for time-dependent systems\cite{Runge1984}, and its linear-response formulation\cite{Casida1995} has become a widely used approach to compute excited-state properties such as the excitation energy and optical spectra. In practical implementations of \gls{lrtddft}, however, the commonly used adiabatic approximation and the reliance on ground-state orbitals lead to several limitations. Rydberg, charge-transfer, and core excitations, which involve a large change of the electron density, tend to be described poorly\cite{Vigneshwaran2025, Maitra2017, Dreuw2003}. In addition, multiple excitations, such as doubly excited states, are absent in standard adiabatic \gls{lrtddft}, and the topology of potential energy surfaces near conical intersections between the ground and excited states is qualitatively incorrect\cite{Maitra2022, Levine2006}. In principle, nonadiabatic extensions of TDDFT provide a general route to overcoming  these limitations, but practical approximations remain an active area of development\cite{Lacombe2023}.

A conceptually simple alternative to TDDFT, which does not rely on a time-dependent framework, is based on variational optimization of the orbitals for the excited state\cite{Herbert2023,Hait2021,Levi2020a}. In practical realizations of this time-\textit{in}dependent approach, the excited states are obtained as stationary points on the electronic energy surface defined by a density functional approximation, corresponding to  solutions of the \gls{ks} equations that lie higher in energy than the ground state. The approach can therefore be viewed as a natural extension of ground-state \gls{ks} \gls{dft} to excited states, in which different electronic states are treated on the same variational footing. The state-specific orbital relaxation is typically found to lead to a more balanced description of excited states with different electronic character, including Rydberg, charge-transfer, and core excitations\cite{Schmerwitz2026,Vigneshwaran2025,Selenius2024,Sigurdarson2023,Hait2021,Ivanov2021,Hait2020core,Barca2018}.

\Gls{oo} density functional calculations of excited states have been explored since the early development of DFT\cite{Ziegler1977, Gunnarsson1976}, but their use has remained far less widespread than that of \gls{lrtddft}. However, in recent years, the field has undergone a sustained revival, driven in part by improved algorithms for locating excited-state stationary points while avoiding variational collapse\cite{Qin2026-mf,Schmerwitz2026,Schmerwitz2023,Carter2020,Hait2020,Levi2020b}. As a result, \gls{oo} density functional methods have evolved into an active and increasingly mature area of research, with methodological developments and applications appearing at an increasing pace\cite{Trushin2026Chemrxiv,Barreiro-Lage2026,Restaino2026arxivTDM,Restaino2026arxivDipole,Malis2026,Yang2026-et,Qin2026-mf,Schmerwitz2026}. 

A note of caution about terminology is needed. Several expressions have appeared in the literature to refer to density functional methods that compute excited states as stationary solutions, the most common being $\Delta$ self-consistent field ($\Delta$SCF). While the term $\Delta$SCF is historically common, it is however broad. It can refer to any calculation in which an energy is obtained as a difference between energy values obtained from separate SCF calculations, and it does not by itself emphasize the defining feature of the approach, namely the state-specific variational optimization of the excited state. Moreover, it does not consider that properties other than the excitation energy can be obtained as well, such as intensities of electronic transitions, dipole moments, and atomic forces, to name a few. In this review, we therefore use the expression \gls{oo} density functional methods or \gls{oo} density functional calculations. More specifically, we focus on fully variational \gls{oo} methods, in which genuine stationary points of the electronic energy surface are sought without imposing additional constraints on the solutions. Related constrained excited-state approaches, which have also seen increasing success lately\cite{Lemke2026,Pham2025,Kussmann2024,Lemke2024-tg,Stella2022,Roychoudhury2020,Ramos2018,Evangelista2013,Gavnholt2008}, are not addressed.

In the present review, we seek to:
\begin{enumerate}
\item Clarify the theoretical foundations of \gls{oo} density functional methods through their connections to formal density functional theories.
\item Illustrate the recent algorithmic advances for orbital optimization of excited states, with particular emphasis on methods designed to locate saddle points on the electronic energy surface.
\item Provide a unified overview of the main \gls{oo} formulations used for open-shell singlet excited states, which have rapidly expanded in recent years.
\item Describe approaches currently used in OO density functional methods to compute transition properties and spectra from generally nonorthogonal, state-specific solutions.
\item Review applications to three representative classes of electronic excitations, namely Rydberg, charge-transfer, and core excited states, as well as calculations of optical spectra, with the aim of assessing where \gls{oo} density functional methods stand with respect to accuracy and range of applicability.
\end{enumerate}

The rapid growth of the field also means that there are other important developments and applications that cannot be covered in a single review. Since \gls{oo} methods are variational, they give access to analytic atomic forces through the Hellmann-Feynman theorem, making simulations of the dynamics of atoms in excited states a prominent example\cite{Birgisson2025, Buchel2024, Mazzeo2023, Vandaele2022, Levi2020pccp, Levi2018}. Closely related is the treatment of nonadiabatic and spin–orbit couplings within \gls{oo} density functional methods, which has recently paved the way to nonadiabatic molecular dynamics simulations based on \gls{oo} states\cite{Vandaele2022, Pradhan2018, Billeter2006}. Another rapidly developing direction is the application of \gls{oo} approaches to electronic excitations in solid-state systems\cite{Sahre2026arxivROKS, Sun2025arxivNVCenter, Ivanov2023, Daga2021, Mackrodt2018}. Although closely connected to the topics discussed in this review, we believe that these developments deserve dedicated reviews and are therefore not covered here.

The article is organized as follows. Section~\ref{sec:theory} discusses the theoretical foundations of excited-state \gls{oo} density functional methods through their connections to exact density functional frameworks. Section~\ref{sec:oo} presents the connection between \gls{oo} excited states and saddle points on the electronic energy surface and the recent advances in algorithms for excited-state orbital optimization. Section~\ref{sec:methods_roks} summarizes the most common OO density functional methods for open-shell singlet excited states. Section~\ref{sec:calculation_spectra} illustrates the approaches used for calculations of transition properties and spectra. Section~\ref{sec:applications} reviews applications to Rydberg, charge-transfer, and core excitations, as well as calculations of optical spectra. Finally, challenges and future perspectives for the development and application of \gls{oo} density functional methods are discussed.

\section{Theory and methods}

\subsection{Theoretical foundations}\label{sec:theory}
\subsubsection{Ground state}
In its original form, \gls{dft} is a ground-state theory. For a given electron density $n(\mathbf r)$, the ground-state energy functional can be written using Levy's constrained search\cite{Levy1979} as
\begin{align}\label{eq:gs_constrained_search}
E^0[n]
&=
\min_{\Psi \to n}
\langle \Psi | \hat T + \hat V_{\mr{ee}} | \Psi \rangle
+
\int v_{\rm ext}(\mathbf r)\, n(\mathbf r)\, d\mathbf r \\ \nonumber 
&= F^0[n] + \int v_{\rm ext}(\mathbf r)\, n(\mathbf r)\, d\mathbf r,
\end{align}
where $\hat T$ and $\hat V_{\mr{ee}}$ are the kinetic and Coulomb electron-electron interaction operators, respectively. Eq.~\ref{eq:gs_constrained_search} defines the universal ground-state functional $F^0[n]$. The ground-state density $n_0(\mathbf r)$ and energy are obtained by minimization of $E^0[n]$. The Hohenberg--Kohn theorem~\cite{Hohenberg1964} states that, for a nondegenerate ground state, the ground-state density determines uniquely the external potential up to an additive constant and, therefore, determines the Hamiltonian and the ground state. 

For a system of non-interacting electrons ($\hat V_\mr{ee}=0$), the universal functional reduces to the non-interacting kinetic energy functional $T^0_\mr{s}[n]$,
\begin{equation}
T^0_\mr{s}[n] =
\min_{\Phi \to n}
\langle \Phi | \hat T | \Phi \rangle.
\end{equation}
The ground-state energy functional can then be written as
\begin{equation}
E^0[n] =
T^0_\mr{s}[n] 
+
E_\mr{H}[n]
+
E^0_{\rm xc}[n]
+ 
\int v_{\rm ext}(\mathbf r)\, n(\mathbf r)\, d\mathbf r,
\end{equation}
where $E_\mr{H}[n]$ is the classical Hartree energy,
\begin{equation}\label{eq:Hartree}
E_\mr{H}[n]
=
\frac{1}{2}
\iint
\frac{n(\mathbf r)n(\mathbf r')}{|\mathbf r-\mathbf r'|}
\, d\mathbf r\, d\mathbf r',
\end{equation}
and $E^0_{\rm xc}[n]$ is the ground-state \gls{xc} functional. In the \gls{ks} approach~\cite{Kohn1965}, the ground-state energy and density are given by minimization over non-interacting wave functions, 
\begin{equation}
E^0=\min_n E^{0}[n]=\min_{\Phi} E^{0}[n_{\Phi}].
\end{equation}
The wave function that minimizes $E^{0}[n_{\Phi}]$ is called the \gls{ks} wave function. In the usual nondegenerate ground-state case, it is a single Slater determinant, although in general situations it is a a spin- and/or spatial-symmetry-adapted linear combination of determinants (a \gls{csf})~\cite{LoosGiarrusso2025,Perdew2003}. When the one-particle density matrix of the non-interacting \gls{ks} state is diagonal, the ground-state density is given by
\begin{equation}
n^0(\mathbf r)=\sum_i f_i^{0}|\psi_i^{0}(\mathbf r)|^2 ,
\end{equation}
where $f_i^{0}$ are the ground-state orbital occupation numbers. The \gls{ks} orbitals are obtained by minimizing the energy with respect to orbital variations, subject to orthonormality. For ordinary ground-state \gls{ks} calculations, the energy is invariant under unitary transformations among orbitals with the same occupation. One may therefore choose a canonical representation in which the stationarity condition takes the form of KS single-particle equations
\begin{equation}
\left[
-\frac{1}{2}\nabla^2
+
v_\mr{H}(\mathbf r)
+
v_{\rm xc}^{0}(\mathbf r)
+
v_{\rm ext}(\mathbf r)
\right]
\psi_i^{0}(\mathbf r)
=
\varepsilon_i^{0}\psi_i^{0}(\mathbf r),
\end{equation}
where the potentials $v_\mr{H}(\mathbf r)$ and $v_{\rm xc}^{0}(\mathbf r)$ are the functional derivatives of the Hartree and \gls{xc} energy with respect to the density.

\subsubsection{Excited states}
While no Hohenberg--Kohn theorem exists for excited states in general~\cite{Gaudoin2004}, it has been shown that it can be extended to the lowest energy state of a given symmetry of any systems of electrons~\cite{Gunnarsson1976}. The corresponding functional is symmetry dependent rather than universal in the ground-state sense.

The absence of a general Hohenberg--Kohn theorem does not preclude the formulation of a general time-independent \gls{dft} for excited states. A general extension of \gls{dft} to excited states has been provided by G\"orling~\cite{Gorling2000, Gorling1999} and recently its role as a formal justification of \gls{oo} density functional calculations has been clarified~\cite{Trushin2026Chemrxiv}. The formalism is based on a stationarity constrained search, in which the minimization in Levy's constrained search, eq.~\ref{eq:gs_constrained_search}, is replaced by the search of a stationary point, providing the density functional associated with a stationary state labeled by a numbering parameter $k$~\footnote{In Görling's formalism, the numbering parameter $k$ does not, in general, coincide with the level of excitation.}
\begin{align}\label{eq:es_constrained_search}
E^k[n]
&=
\underset{\Psi \to n} {\mathrm{stat}\,}
\langle \Psi | \hat T + \hat V_{\mr{ee}} | \Psi \rangle
+
\int v_{\rm ext}(\mathbf r)\, n(\mathbf r)\, d\mathbf r \\ \nonumber 
&= F^k[n] + \int v_{\rm ext}(\mathbf r)\, n(\mathbf r)\, d\mathbf r.
\end{align}
The exact density and energy of the stationary state associated with label $k$ are obtained from a stationarity condition on $E^k[n]$. When the chosen stationary solution is the absolute minimum, this reduces to the ground state~\cite{LoosGiarrusso2025, Gorling1999}. Görling introduced a density theorem, according to which the density of any eigenstate of an electron system determines the external potential, and hence the Hamiltonian and all properties of the electronic system~\cite{Trushin2026Chemrxiv, Gorling1999}. The density theorem is general, as it is valid for both ground- and excited-state densities. The Hohenberg–Kohn theorem is recovered as the special case in which the eigenstate is the ground state. 

The \gls{ks} approach has also be extended to excited states by Görling~\cite{Trushin2026Chemrxiv, Gorling1999}. The density theorem holds for physical interacting as well as non-interacting electron systems. However, there exist several eigenstates of different physical electron systems and several eigenstates of different KS systems with the same electron density. A generalized adiabatic connection~\cite{Trushin2026Chemrxiv, Gorling1999} provides the connection between an interacting stationary state and a particular non-interacting \gls{ks} stationary state with the same density and the same numbering parameter $k$. The energy functional for state $k$ can then be written as
\begin{equation}
E^k[n]
=
T^k_\mr{s}[n]
+
E_\mr{H}[n]
+
E^k_{{\rm xc}}[n]
+
\int v_{\rm ext}(\mathbf r)\, n(\mathbf r)\, d\mathbf r
,
\end{equation}
where $T^k_\mr{s}[n] = \underset{\Phi \to n} {\mathrm{stat}\,} \langle \Phi | \hat T | \Phi \rangle$ is the non-interacting kinetic energy functional, and the xc functional is a functional of the numbering parameter $k$. The density and energy of the stationary state with label $k$ follow from the stationarity of the corresponding energy functional,
\begin{equation}
E^k=\underset{n}{\mathrm{stat}}\, E^{k}[n]
=
\underset{\Phi}{\mathrm{stat}}\, E^{k}[n_{\Phi}] .
\end{equation}
Here, the \gls{ks} state is not required to be the ground state of the non-interacting Hamiltonian. As for the ground state, the \gls{ks} wave function $\Phi^k$ may be a single Slater determinant, but in general it is a \gls{csf}\cite{LoosGiarrusso2025, Gorling2000, Gunnarsson1976}.  When the one-particle density matrix of the non-interacting \gls{ks} state is diagonal, the density associated with state $k$ can be written as
\begin{equation}
n^k(\mathbf r)=\sum_i f_i^k |\psi_i^k(\mathbf r)|^2,
\end{equation}
where the occupation numbers $f_i^k$ are such that a set of orbitals different from the lowest energy ones are occupied, i.e. the orbital occupation is nonaufbau~\cite{Gorling1999}.
The \gls{ks} orbitals are obtained from the stationarity of $E^k[n_\Phi]$ with respect to orbital variations, subject to orthonormality. As for the ground state, when the energy is invariant under unitary transformation among equally occupied orbitals, this is equivalent to solving single-particle \gls{ks} equations, 
\begin{equation}
\left[
-\frac{1}{2}\nabla^2
+
v_\mr{H}(\mathbf r)
+
v_{\rm xc}^k(\mathbf r)
+
v_{\rm ext}(\mathbf r)
\right]
\psi_i^k(\mathbf r)
=
\varepsilon_i^k\psi_i^k(\mathbf r).
\end{equation}

Görling noted that in the excited-state KS formalism, the kinetic and \gls{xc} contributions have the same formal expressions as for the ground state when written in terms of the associated KS and interacting wave functions~\cite{Trushin2026Chemrxiv, Gorling1999}. This provides a formal rationale for using the same approximate functionals for both ground and excited states. A related formal foundation of OO density functional methods for excited states based on potential functionals has recently been proposed by Yang and Ayers \cite{Yang2024}, who use additional variables, such as the potential, of a non-interacting reference system rather than the excited-state density alone. Its relation to Görling’s density theorem and generalized adiabatic connection KS formalism is discussed in reference \citenum{Trushin2026Chemrxiv}.

Practical \gls*{oo} density functional approaches use ordinary functionals originally developed for ground-state calculations. The energy of state $k$ is then obtained as a stationary point of the approximate ground-state functional evaluated with a non-interacting state $\Phi^k$. In this approximation, the explicit dependence of the exact \gls{xc} functional on the numbering parameter $k$ present in the exact excited-state theory of Görling is neglected. Nevertheless, a state dependence enters the calculation through the nonaufbau orbital occupation and the corresponding state-specific optimized orbitals. 

Also partially rationalizing the use of a ground-state functional to compute excited states, earlier Perdew and Levy~\cite{Perdew1985} showed that every stationary point of the exact ground-state energy functional $E^0[n]$ corresponds to the density of a stationary state of the interacting system. However, not every excited-state density appears as a stationary point of $E^0[n]$. 

Ayers, Levy, and Nagy~\cite{Ayers2012} later formulated a time-independent excited-state \gls{dft} specifically for the class of Coulomb systems, which includes atoms and molecules. This theory exploits the property that a Coulomb density determines both the external potential, through the conventional cusp conditions~\cite{ParrYangBook}, and the excitation level, because the asymptotic decay of the electron density of a state is defined by the ionization potential of that state~\cite{Ayers2012}. On that basis, stationarity principles with a Coulomb functional $F_{\mathrm{Coul}}[n]$ as well as a state-dependent variant $F^k_{\mathrm{Coul}}[n]$, where $k$ represents the excitation level, were proposed. Since this construction is exact only within the class of Coulomb external potentials, the functional is best viewed as subuniversal. While it was highlighted that $F_{\mathrm{Coul}}[n]$ may be a ``jagged, discontinuous functional'', this formulation was later extended to the \gls{ks} approach by defining the corresponding non-interacting kinetic-energy functional and deriving single-particle \gls{ks} equations for the excited-state density~\cite{Ayers2015}.

Following the formulation of such time-independent density functional theories for excited states, efforts have more recently focused on the development of excited-state-specific functionals~\cite{Loos2025, LoosGiarrusso2025, Hemanadhan2014}. For example, recently Loos introduced an excited-state uniform electron gas characterized by a gap at the Fermi surface and proposed including an additional variable into the functional measuring the degree of excitation~\cite{Loos2025}. Despite promising developments, however, most practical \gls*{oo} density functional approaches still rely on ordinary ground-state \gls{xc} functionals and seek excited states as stationary nonaufbau solutions of the corresponding approximate \gls{ks} equations.

Recent work has also connected practical \gls*{oo} density functional methods to ensemble \gls{dft}~\cite{Oliveira1988EnsembleDFTII, Gross1988EnsembleDFTI, Gross1988RayleighRitz, Theophilou1979}. In particular, Gould and Fromager~\cite{Fromager2025, Gould2025pra} have shown  that the energy of an individual ground or excited state can be expressed as a functional of the ensemble density $n$,
\begin{equation}
E^{\xi,k}[n]
=
T_\mr{s}^{\xi,k}[n]
+
E_{{\rm Hxc}}^{\xi,k}[n]
+
\int v_{\rm ext}(\mathbf r)\,n^{\xi,k}[n](\mathbf r)\,d\mathbf r,
\end{equation}
where $\xi$ is a set of ensemble weights, and $E_{{\rm Hxc}}^{\xi,k}$ is component $k$ of the ensemble Hartree-exchange-correlation (Hxc) functional. Each individual state fulfills a stationarity condition with respect to the ensemble density
\begin{equation}
E^k=\underset{n}{\mathrm{stat}}\, E^{\xi,k}[n].
\end{equation}
Within this context, Fromager has shown~\cite{Fromager2025} that practical \gls*{oo} density functional approaches are recovered when the ensemble Hxc functional is approximated by recycling conventional ground-state xc functionals and so-called density-driven correlations are ignored~\cite{Dupuy2026, Gould2025pra}. This provides an alternative, ensemble-derived theoretical underpinning for \gls*{oo} approaches.

Formal links between \gls*{oo} density functional calculations of excited states and TDDFT have also been established~\cite{Park2015, Kowalczyk2011, Ziegler2009}. Kowalczyk, Yost, and Van Voorhis~\cite{Kowalczyk2011} have shown that, within the adiabatic approximation, \gls*{oo} solutions correspond to stationary densities of the time-dependent \gls{ks} equations, satisfying $\dot{n}({\bf{r}},t)=0$. This condition is only necessary, not sufficient, since not all stationary densities correspond to true eigenstates. For example, stationary densities can be provided by states that break the spin symmetry (the problem of spin mixing in the case of open-shell singlet excited states and methods to deal with it are described in section~\ref{sec:methods_roks}). Ziegler and co-workers~\cite{Park2015, Ziegler2009} have given a different perspective by showing that adiabatic TDDFT can be derived from a constrained variational treatment of the \gls{ks} energy. In this perspective, practical implementations of TDDFT can be considered second-order variational methods, while \gls*{oo} approaches are a higher-order extension.

Overall, practical \gls*{oo} density functional approaches for excited states find partial justification within several theoretical frameworks, including state-specific and ensemble time-independent \glspl{dft}, as well as adiabatic TDDFT. Their working equations are obtained by reusing ordinary ground-state \gls{xc} functionals~\cite{Trushin2026Chemrxiv, Dupuy2026, Fromager2025, Gorling1999}, an approximation that is sometimes considered rather crude from the perspective of the exact theories. Yet, \gls*{oo} methods have proven to be highly successful in practice, and often outperform standard TDDFT, as will be illustrated in section ~\ref{sec:applications}.

\subsection{Excited-state orbital optimization}\label{sec:oo}
Orbital-optimized excited states are more challenging to obtain than the ground state, because unlike the latter, they are generally not minima of the surface given by the variation of the electronic energy as a function of the electronic degrees of freedom. Instead, they usually correspond to saddle points~\cite{Schmerwitz2023, Hait2021}. This can be seen by considering a unitary rotation of the molecular orbitals,
\begin{equation}\label{eq:unitary_transformation}
\boldsymbol{\psi}^\prime_k
=
\boldsymbol{\psi}_k\mathbf U ,
\end{equation}
where $\boldsymbol{\psi}_k$ is a vector of orthonormal molecular orbitals (occupied and unoccupied). The unitary matrix can be parametrized as an exponential of an anti-Hermitian matrix $\kappa$ of orbital rotation angles,
\begin{equation}\label{eq:exp_transformation}
\mathbf U
=
e^{\boldsymbol{\kappa}},
\qquad
\boldsymbol{\kappa}^{\dagger}
=
-\boldsymbol{\kappa}.
\end{equation}
At a stationary solution $k$, the curvature of the electronic energy with respect to orbital rotations is described by the electronic Hessian. For a \gls{ks} functional in the case of real, canonical orbitals, the elements of the Hessian are~\cite{Ziegler2008, Bauernschmitt1996}
\begin{align}\label{eq:hessian}
\frac{\partial^2 E^{k}}
{\partial \kappa_{ij}\partial \kappa_{lm}} = 
&-2(\varepsilon^k_i-\varepsilon^k_j)(f_i^{k}-f_j^{k})\delta_{il}\delta_{jm}
\\ \nonumber
&+ 4(f_i^{k}-f_j^{k})(f_l^{k}-f_m^{k})
(ij|f_{\mathrm{Hxc}}|lm),
\end{align}
where the spin index has been omitted for compactness. The kernel matrix element consists of Hartree and \gls{xc} parts:
\begin{equation}
(ij|f_{\mathrm{Hxc}}|lm)
=
(ij|lm)
+
(ij|f_{\mathrm{xc}}|lm),
\end{equation}
with
\begin{equation}
(ij|lm)
=
\iint
\psi^k_i(\mathbf r)\psi^k_j(\mathbf r)
\frac{1}{|\mathbf r-\mathbf r'|}
\psi^k_l(\mathbf r')\psi^k_m(\mathbf r')
\,d\mathbf r\,d\mathbf r'
\end{equation}
and
\begin{equation}
(ij|f_{\mathrm{xc}}|lm)
=
\iint
\psi^k_i(\mathbf r)\psi^k_j(\mathbf r)
\frac{\delta^2 E_{\mathrm{xc}}}
{\delta n(\mathbf r) \delta n(\mathbf r')}
\psi^k_l(\mathbf r')\psi^k_m(\mathbf r')
\,d\mathbf r\,d\mathbf r',
\end{equation}
where $f_{\mathrm{xc}}$ is referred to as the \gls{xc} kernel. The dominant contribution often comes from the one-electron diagonal terms~\cite{Levi2020b}, which involve orbital energy differences. Excited-state solutions typically correspond to nonaufbau occupations, where some higher-energy orbitals are occupied while lower-energy orbitals are empty. Then, according to eq.~\ref{eq:hessian}, rotations connecting higher-energy occupied and lower-energy unoccupied orbitals give negative curvatures. Thus, orbital-optimized excited states are naturally associated with saddle points of the electronic energy surface.

\begin{figure}
\centering
\includegraphics[width = 0.9\linewidth]{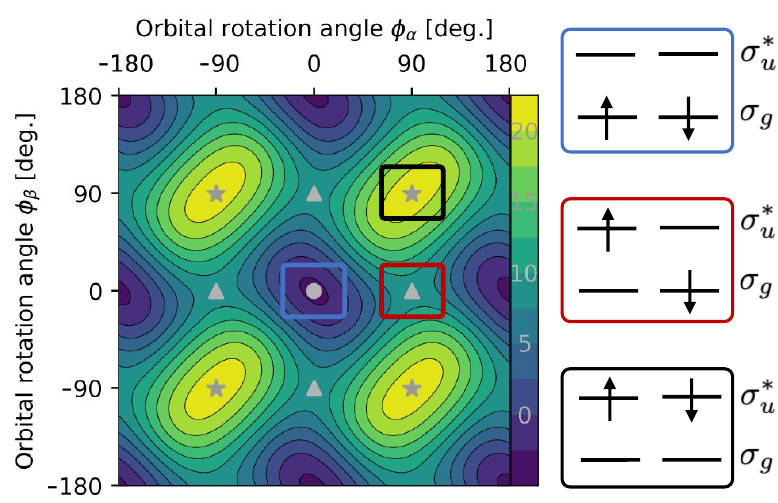}
\caption{Electronic energy surface of H$_2$ from \gls{oo} unrestricted calculations with the PBE functional in a minimal basis set, shown as a function of the orbital rotation angles $\phi_\alpha$ and $\phi_\beta$ mixing the $\sigma_g$ and $\sigma_u^*$ orbitals in the two spin channels. The minimum corresponds to the ground-state solution $\sigma_g^2\sigma_u^{*0}$, while the first- and second-order saddle points correspond to the singly excited open-shell $\sigma_g^1\sigma_u^{*1}$ and doubly excited $\sigma_g^0\sigma_u^{*2}$ solutions, respectively. The degenerate, equivalent solutions are related by a sign change of the orbitals. Adapted with permission from Y. L. A. Schmerwitz, G. Levi and H. Jónsson ``Calculations of Excited Electronic States by Converging on Saddle Points Using Generalized Mode Following'', J. Chem. Theory Comput. \textbf{19}, 3634–3651 (2023). Copyright 2023, American Chemical Society.}
\label{fig:h2_saddle_points}
\end{figure}

Figure~\ref{fig:h2_saddle_points} illustrates the connection between excited-state solutions and saddle points for the H$_2$ molecule, described with \gls{oo} spin-unrestricted \gls{ks} calculations in a minimal basis set. 
The electronic energy surface is shown as a function of the only two independent orbital rotation angles available in this basis, $\phi_\alpha$ and $\phi_\beta$, which mix the bonding and antibonding orbitals in the $\alpha$ and $\beta$ spin channels, respectively. For each point on the surface, the spin orbitals are obtained from the ground-state bonding and antibonding orbitals, $\sigma_g$ and $\sigma_u^*$, through the unitary transformation
\begin{equation*}
    \begin{pmatrix}
        \vspace{1pt}
        \psi_{1}^{\alpha}\\
        \vspace{1pt}
        \psi_{0}^{\alpha}\\
        \vspace{1pt}
        \psi_{1}^{\beta}\\
        \vspace{1pt}
        \psi_{0}^{\beta}
    \end{pmatrix}
    =
    \begin{pmatrix}
            \cos{\phi_{\alpha}} & \sin{\phi_{\alpha}} & 0 & 0\\
            -\sin{\phi_{\alpha}} & \cos{\phi_{\alpha}} & 0 & 0\\
            0 & 0 & \cos{\phi_{\beta}} & \sin{\phi_{\beta}}\\
            0 & 0 & -\sin{\phi_{\beta}} & \cos{\phi_{\beta}} 
    \end{pmatrix}
    \begin{pmatrix}
        \vspace{1pt}
        \sigma_{g}\\
        \vspace{1pt}
        \sigma^{*}_{u}\\
        \vspace{1pt}
        \sigma_{g}\\
        \vspace{1pt}
        \sigma^{*}_{u}
    \end{pmatrix}\,.
\end{equation*}
At $\phi_\alpha=\phi_\beta=0$, the orbitals correspond to the ground-state solution, which has configuration $\sigma_g^2\sigma_u^{*0}$ and is a minimum on the electronic energy surface. The surface also contains saddle points corresponding to excited-state solutions. The first order saddle points correspond to the open-shell, spin-mixed excited-state solution $\sigma_g^1 \sigma_u^{*1}$ obtained by a $\pm 90^\circ$ rotation in one spin channel. The second order saddle points correspond to a doubly excited state solution $\sigma_g^0 \sigma_u^{*2}$ obtained by a $\pm 90^\circ$ rotation in both spin channels.

The structure of the Hessian suggests that the saddle-point order tends to increase with the excitation level. This trend, however, is not strictly monotonic for nonlinear variational calculations, as proven for \gls*{oo} Hartree-Fock and complete active space self-consistent field (CASSCF) calculations by Burton~\cite{Marie2023, Burton2022} (similar studies for \gls*{oo} density functional calculations are, to our knowledge, still lacking). This behavior contrasts with that of linear variational methods, such as \gls{ci}. There, the linear parametrization of the wave function provides a simpler structure of the electronic Hessian, with eigenvalues around a stationary state $k$ given by $2(E^n-E^k)$~\cite{MolecularElectronicStructureTheory}, where $n$ labels the other states. Therefore, every state below the state $k$ contributes one negative curvature direction, and as a result, the first excited state is a first-order saddle point, the second excited state is a second-order saddle point, and so forth~\cite{Burton2022, MolecularElectronicStructureTheory}. 

\subsubsection{Choice of initial guess}\label{sec:initial_guess}
\begin{figure*}[ht!]
\centering
\includegraphics[width = 0.7\textwidth]{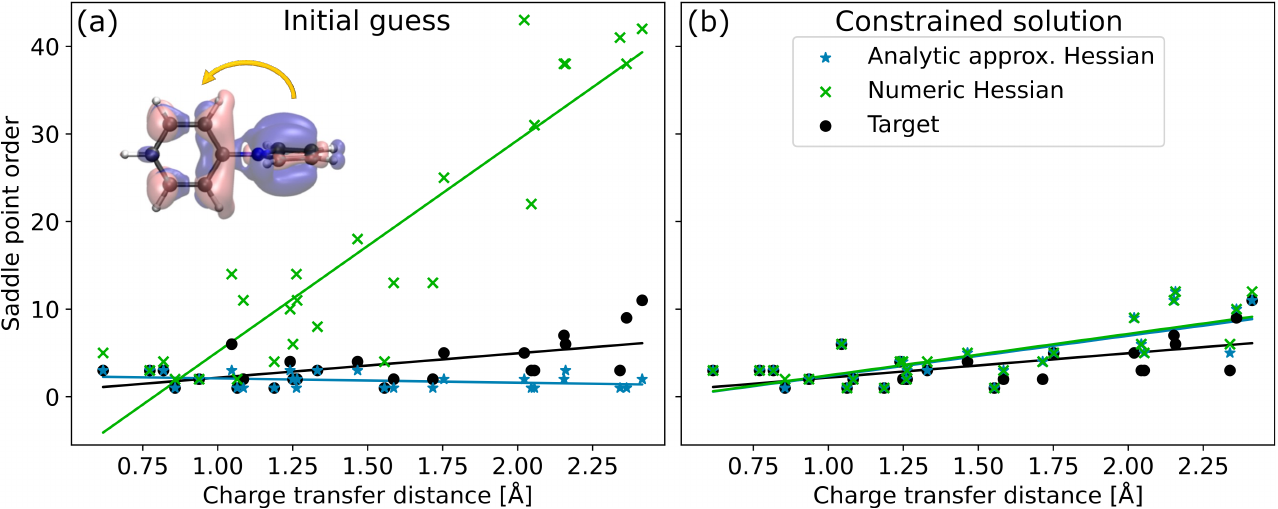}
\caption{Estimated saddle point order of intramolecular charge-transfer excited states of organic molecules as a function of charge transfer distance for (a) an initial guess of ground-state orbitals with changed occupations and (b) after a constrained optimization in which the hole and particle orbitals are frozen. Black dots represent the saddle point order of the target solutions. Blue stars represent the saddle point order obtained from an analytic Hessian approximation including only the diagonal terms depending on the orbital energy differences (see eq. \ref{eq:hessian}), and green crosses represent the saddle point order obtained numerically. The calculations use the PBE functional and are spin-unrestricted. The inset shows one of the molecules included in the study, $N$-Phenylpyrrole, together with the electron density difference between the ground state and an excited state involving charge transfer from the pyrrole to the phenyl group. Adapted from Y. L. A. Schmerwitz, E. Selenius, and G. Levi, ``Freeze-and-Release Direct Optimization Method for Variational Calculations of Excited Electronic States'', J. Chem. Theory Comput. \textbf{22}, 3571–3584 (2026). Published by American Chemical Society under the Creative Commons Attribution 4.0 International License (CC BY 4.0).}
\label{fig:initial_guess_spo}
\end{figure*} 
The choice of the initial guess is considerably more critical in \gls*{oo} density functional calculations of excited states than in ground-state calculations. For ground-state calculations, the minimum energy principle provides a simple prescription: One must minimize the energy to obtain the global minimum, and any reasonable initial density lying within its basin of attraction can be used for this task. In \gls*{oo} calculations, by contrast, the target is a higher-energy nonaufbau stationary solution, typically a saddle point. Therefore, when targeting a specific excited state, the initial guess must encode the identity of the target excited state. Moreover, excited states are often close in energy\cite{Dreuw2026} and a small change in the initial guess may direct the optimization toward a different stationary solution. 

In practice, the construction of such an initial guess often relies on prior information about the target excitation. A possible route is to perform a preliminary \gls{lrtddft} calculation and inspect the eigenvectors obtained from Casida's equation~\cite{Qin2026-mf, Selenius2024, Sinyavskiy2025}. Typically, the dominant electron–hole contribution for a given state is used to define a nonaufbau occupation pattern. Then, starting from the ground-state \gls{ks} canonical orbitals, an electron is promoted from the hole orbital to the particle orbital. However, several works have shown that such initial guess of canonical ground-state orbitals is often not optimal. This is particularly clear for charge-transfer and core-level excitations, where orbital relaxation is large~\cite{Qin2026-mf, Schmerwitz2026, Bogo2025, Selenius2024, Bogo2024, Hait2021, Hait2020core}. Figure~\ref{fig:initial_guess_spo} illustrates this point by comparing the saddle point order evaluated at an initial guess of ground-state orbitals with changed occupations with that of the final solution in \gls{oo} unrestricted KS calculations of charge-transfer states in organic molecules~\cite{Schmerwitz2026}. At the initial guess of ground-state canonical orbitals, the numerical Hessian significantly overestimates the saddle-point order, whereas an analytical approximation including only the diagonal terms depending on the orbital energy differences in eq.~\ref{eq:hessian} underestimates it. This indicates that the electronic energy surface around such initial guess is highly rugged~\cite{Schmerwitz2026}. After a constrained optimization in which the orbitals directly involved in the excitation are kept fixed while the remaining orbitals are relaxed (see the next section), the estimated saddle-point order is much closer to that of the final solution~\cite{Schmerwitz2026}. A similar conclusion was reached by Stein and co-workers~\cite{Bogo2025}, who refer to the constrained optimization as a way to generate an improved guess for \gls{oo} calculations of charge-transfer excited states. 

\begin{figure}
\centering
\includegraphics[width = 0.9\linewidth]{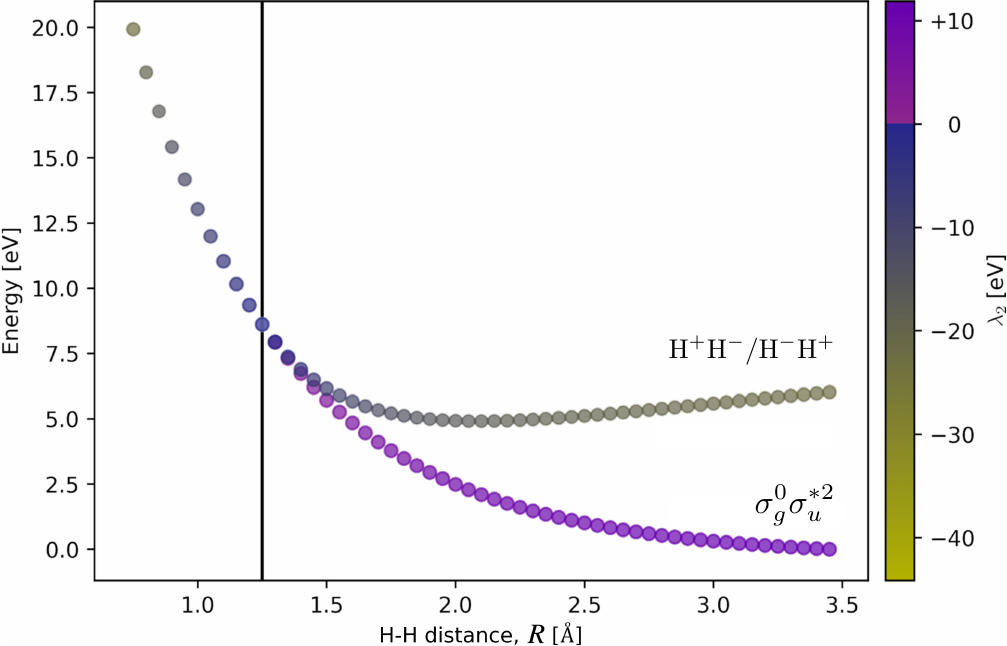}
\caption{Energy of the lowest doubly excited state of H$_2$ as a function of the bond length obtained in orbital-optimized spin-unrestricted calculations using the PBE functional and a minimal basis set. The vertical line indicates the point where two solutions emerge. The points are colored according to the value of the second lowest eigenvalue of the electronic Hessian. The upper solution is ionic, breaks the spatial symmetry, and corresponds to a second-order saddle point. The lower solution preserves the symmetry and corresponds to a first-order saddle point. Adapted from Y. L. A. Schmerwitz, N. U. Ollé, G. Levi, and H. Jónsson, ``Saddle Point Search Algorithms for Variational Density Functional Calculations of Excited Electronic States with Self-Interaction Correction''. PASC ’24 Proc. Platf. Adv. Sci. Comput. Conf. \textbf{19}, 1-11 (2024). Published by Association for Computing Machinery under the Creative Commons Attribution 4.0 International License (CC BY 4.0).}
\label{fig:h2_curves_sp_order}
\end{figure} 
The choice of the initial guess is a particularly important consideration when multiple mean-field solutions are associated with the same excitation. Figure~\ref{fig:h2_curves_sp_order} shows the case of \gls{oo} unrestricted KS calculations of the lowest doubly excited state of H$_2$ along the curve given by the energy as a function of the bond length~\cite{Schmerwitz2024, Schmerwitz2023}. Close to the ground-state geometry, there exists a single solution with configuration $\sigma_g^0 \sigma_u^{*2}$ corresponding to a second-order saddle point (see also Figure \ref{fig:h2_saddle_points}). At longer bond lengths, after a point of symmetry-breaking onset, two solutions exist. The lower-energy solution is a first-order saddle point while the upper-energy solution is symmetry-broken and corresponds to a second-order saddle point. The lower-energy branch corresponds to a double excitation that preserves the spatial symmetry, whereas the upper branch corresponds to an ionic solution that breaks the spatial symmetry but is in qualitative agreement with full CI results. Starting from (symmetry-adapted) canonical ground-state orbitals naturally biases the calculation toward the symmetry-preserving branch. Following the branch with a qualitative correct energy curve requires the algorithm to break the symmetry in the initial guess in an appropriate way. An analogous issue occurs in \gls*{oo} density functional calculations of core-level excitations, as reported by several authors~\cite{Qin2026-mf, Hait2021, Hait2020core, Besley2009}. There, often the desired state corresponds to a localized core hole. If several equivalent atoms are present, a symmetry-adapted guess, as the orbitals of a ground-state calculations, can delocalize the hole over several atoms, which exacerbates delocalization errors inherent in the \gls{xc} functional. Therefore, the symmetry of the initial guess needs to be broken to localize the hole on an atomic site.

Despite these limitations, most current calculations still begin from canonical ground-state orbitals with manually changed occupations. A more robust approach may be to use orbitals that are already adapted to the transition, such as natural (transition) orbitals from a preliminary TDDFT calculation. An initial guesses of natural orbitals has been noted to improve convergence in state-specific excited-state HF calculations\cite{Kossoski2022}, and recently an approach has been proposed that uses transition orbitals from a calculation within the \gls{tda} to generate an initial guess for \gls*{oo} density functional calculations of core excited states~\cite{Qin2026-mf}. More systematic benchmarks are needed to determine whether natural orbitals generally outperform common initial guesses of ground-state canonical orbitals.

Finally, most practical approaches are state-specific, in that they use prior information to converge one target excited state. An alternative strategy is to simultaneously search for multiple SCF solutions (semi-)globally. For example, one may rotate occupied and unoccupied ground-state orbitals within an energy window to generate several nonaufbau guesses, or use global optimization strategies to locate multiple stationary solutions. Basin hopping algorithms have been recently proposed for global exploration of SCF solutions in the context of Hartree-Fock~\cite{Dong2020}. Such approaches would offer a route toward obtaining \gls*{oo} excited states without prior information and finding excited-state solutions far from initial guesses of ground-state orbitals, which may otherwise be missed.

\subsubsection{Optimization algorithms} \label{sec:oo_algorithms}
The convergence of \gls{oo} excited-state calculations depends critically on the algorithm used to solve the SCF problem. Since the \gls{oo} excited states typically correspond to saddle points, the main risk is convergence to a lower-energy solution, the so-called variational collapse. Indeed, this has long been one of the main obstacles to the widespread application of fully variational \gls{oo} density functional approaches, and it still in part limits their applicability.

One of the first and still most commonly used strategies for reducing the risk of variational collapse is the \gls{mom}~\cite{Sinyavskiy2025,Corzo2022,Macetti2021,Barca2018,Gilbert2008,Cheng2008}. MOM is not by itself an optimization algorithm. Rather, it is a prescription for choosing the orbital occupations during an SCF calculation. In the original, 2008 formulation of MOM by Gill and co-workers\cite{Barca2018, Gilbert2008}\footnote{A similar maximum overlap criterium was also used in an application to Rydberg states by Van Voorhis and co-workers in the same year~\cite{Cheng2008}.}, at each SCF iteration $n$, the occupied orbitals are chosen as those having the largest projections onto the space of a set of reference orbitals,
\begin{equation}\label{eq:mom}
\omega^{(n)}_j=\sqrt{\sum_{i\in{\rm occ}} |S^{(n)}_{ij}|^2},
\end{equation}
where $S_{ij}^{(n)}$ is the overlap between orbital $j$ at the iteration $n$ with orbital $i$ of the reference set, $S_{ij}^{(n)} = \langle \psi^{(n)}_{{\rm ref},i} | \psi_j^{(n)} \rangle$. In the first version of MOM\cite{Gilbert2008}, the reference orbitals were chosen as the occupied orbitals of the previous SCF iteration~\cite{Gilbert2008}, i.e. $\psi^{(n)}_{{\rm ref},i} \equiv \psi^{(n-1)}_{i}$. However, this choice can make the calculation drift away from the target excited-state solution gradually, eventually leading to collapse to an undesired solution\cite{Carter2020, Hait2020, Barca2018, Mewes2014}. The presently most commonly used MOM algorithm, often referred to as the initial maximum overlap method (IMOM), keeps the reference orbitals fixed to the initial excited-state guess~\cite{Barca2018}, i.e. $\psi^{(n)}_{{\rm ref},i} \equiv \psi^{(0)}_{i}$. In practice, IMOM changes the occupation numbers when the distance from the initial guess becomes large, i.e. when the calculation drifts too much away from the initial guess. The change of occupations leads to a jump on the electronic energy surface. After such a jump, one hopes that the following SCF iterations converge to a stationary solution close to the initial guess. Thus, IMOM does not guarantee convergence. Ultimately, convergence is determined by the underlying orbital-optimization algorithm.

\begin{figure*}[ht!]
\centering
\includegraphics[width = 0.75\textwidth]{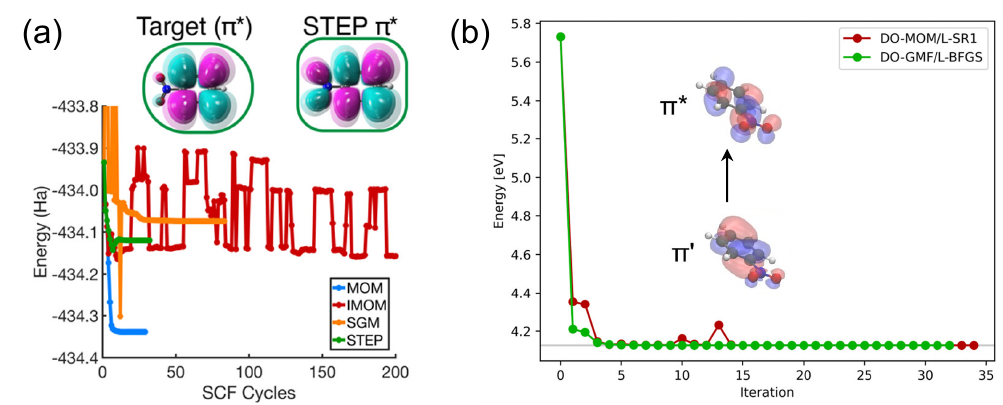}
\caption{Comparison of optimization algorithms for calculations of charge-transfer excited states of nitrobenzene. 
(a) Calculations of an excitation from a $\pi$ orbital on the nitro group to a $\pi^*$ orbital on the phenyl ring. Calculations with the maximum overlap method (MOM) and initial MOM collapse to the ground state and oscillate without convergence, respectively. A calculation with square gradient minimization (SGM) converges to a spurious charge-delocalized solution. The state targeted energy projection (STEP) method, which applies level shifting to the unoccupied orbitals, converges to the target charge-transfer solution. MOM, IMOM and STEP calculations use an SCF algorithm based on eigendeceomposition of the Hamiltonian matrix, while SGM uses a direct minimization approach. All calculations use Hartree-Fock and are spin-unrestricted. Failures are also observed for orbital-optimized unrestricted Kohn--Sham calculations of the same state. 
(b) Direct optimization (DO) calculations of an excitation from a $\pi$ orbital on the phenyl ring to a $\pi^*$ orbital on the nitro group. The DO-MOM calculation uses the L-SR1 quasi-Newton algorithm, while DO with generalized mode following (GMF) targets a fourth-order saddle point. Both methods converge to the target charge-transfer solution. 
Panel (a) adapted with permission from K. Carter-Fenk and J. M. Herbert, ``State-Targeted Energy Projection: A Simple and Robust Approach to Orbital Relaxation of Non-Aufbau Self-Consistent Field Solutions'', J. Chem. Theory Comput. \textbf{16}, 5067--5082 (2020). Copyright 2020, American Chemical Society. 
Panel (b) adapted with permission from Y. L. A. Schmerwitz, G. Levi, and H. J\'onsson, ``Calculations of Excited Electronic States by Converging on Saddle Points Using Generalized Mode Following'', J. Chem. Theory Comput. \textbf{19}, 3634--3651 (2023). Copyright 2023, American Chemical Society.}
\label{fig:nitrobenzene_convergence}
\end{figure*} 
Typically, \gls{mom}-based calculations use conventional SCF algorithms based on Hamiltonian diagonalization and convergence acceleration by direct inversion in the iterative subspace (DIIS)\cite{Pulay1980}. Such algorithms are highly effective for the ground state, which is a minimum on the electronic energy surface, but they are not designed specifically for saddle points. Moreover, DIIS calculations can show erratic convergence when, e.g., unequally occupied orbitals are nearly degenerate, or the energetic order and character of orbitals change significantly during the optimization. Charge-transfer excitations provide an example because there the orbitals can undergo large relaxation. A prototypical case are charge-transfer excitations in nitrobenzene. Figure ~\ref{fig:nitrobenzene_convergence} illustrates that for a calculation of an excited state of nitrobenzene involving excitation of an electron from an orbital localized on the nitro group to an orbital localized on the benzene ring, DIIS combined with MOM collapses to the ground state, while DIIS with IMOM oscillates without converging for $\sim$200 iterations~\cite{Carter2020}. 

A more recent but also popular approach is the state-targeted energy projection (STEP) method of Carter-Fenk and Herbert~\cite{Carter2020}. There, the Hamiltonian matrix is modified by including a projector that raises the energy of the virtual orbitals. Such level shifting restrains occupied--virtual orbital rotations reducing the risk of variational collapse. In the original formulation, it is used with a standard diagonalization-based SCF algorithm. STEP has been shown to converge the problematic charge-transfer states of nitrobenzene for which DIIS-MOM collapses to the ground state and DIIS-IMOM shows erratic convergence behavior~\cite{Carter2020} (see Figure~\ref{fig:nitrobenzene_convergence}). While the computational cost per iteration is the same as an ordinary SCF calculation, STEP is reliant on the quality of the initial guess and might require a large number of iterations due to the restrictions imposed on the orbital rotations.

Some of the limitations of diagonalization-based algorithms such as DIIS can be overcome using direct optimization (DO) methods, which instead of solving the KS eigenvalue equations, directly seek a unitary transformation that makes the energy stationary with respect to variations of the orbitals that preserve orthonormality. Typically, the unitary transformation of the orbitals is parametrized as the exponential of an anti-Hermitian matrix, $\boldsymbol{\kappa}$ (see eq. \ref{eq:exp_transformation}). Thus, based on eqs. \ref{eq:unitary_transformation} and \ref{eq:exp_transformation}, in general the optimization problem consists of making the energy stationary with respect to $\boldsymbol{\kappa}$ and minimizing it with respect to $\boldsymbol{\psi}$, as shown by Ivanov et al.\cite{Ivanov2021},
\begin{equation}\label{eq:do_general}
 \underset{\boldsymbol{\psi}^\prime}{\mathrm{stat}\,} E[\boldsymbol{\psi}^\prime]
 =
 \underset{\boldsymbol{\psi}}{\mathrm{min}\,}
 \underset{\boldsymbol{\kappa}}{\mathrm{stat}\,}
 E[\boldsymbol{\psi}e^{\boldsymbol{\kappa}}]\,.
\end{equation}
Most commonly, \gls{oo} density functional calculations are performed in the \gls*{lcao} representation, in which the orbitals are expanded as
\begin{equation}
    \boldsymbol{\psi}=\boldsymbol{\chi}\mathbf C,
\end{equation}
where $\boldsymbol{\chi}$ denotes a vector of basis set functions and $\mathbf C$ a matrix of coefficients. Since the basis functions are fixed, the variational problem reduces to the optimization of the independent orbital rotation parameters,
\begin{equation}\label{eq:do_lcao}
    \underset{\boldsymbol{\psi}^\prime}{\mathrm{stat}}\,
    E[\boldsymbol{\psi}^\prime]
    =
    \underset{\boldsymbol{\kappa}}{\mathrm{stat}}\,
    E[\boldsymbol{\chi}\mathbf Ce^{\boldsymbol{\kappa}}],
\end{equation}
yielding the optimal coefficients
\begin{equation}
    \mathbf C^\prime=\mathbf Ce^{\boldsymbol{\kappa}}.
\end{equation}
In calculations of the ground state, the search for a stationary point reduces to a minimization. It has long been known that direct minimization methods are more robust than diagonalization-based SCF algorithms, especially when nearly degenerate unequally occupied orbitals are involved.\cite{Lehtola2020, Voorhis2002, Head-Gordon1988} 

For excited states, the direct optimization of the energy involves finding a saddle point in the space of anti-Hermitian matrices, which is a more complicated task compared to minimization. The \gls{sgm} algorithm by Hait and Head-Gordon~\cite{Hait2020} recasts the problem into a minimization by  using the square of the gradient of the energy with respect to orbital rotations, $\left|\nabla_{\boldsymbol{\kappa}} E\right|^2$, rather than the energy as the objective function. All stationary points, including saddle points, of the electronic energy are minima of the squared gradient objective function. So, the approach can be used to obtain \gls{oo} excited states through, e.g., efficient quasi-Newton methods for unconstrained minimization, such as the Broyden–Fletcher–Goldfarb–Shanno (BFGS) algorithm. \gls{sgm} has been shown to be robust for \gls{oo} excited-state calculations where DIIS- and MOM-based algorithms struggle~\cite{Bogo2024,Hait2020}. However, the method is more expensive than ordinary ground-state orbital optimization because the gradient of the squared gradient must be evaluated. Moreover, minima on the squared gradient landscape can be connected by unphysical stationary points with small barriers\cite{Burton2022, Cuzzocrea2020}, which may lead to convergence to spurious solutions. For example, for the charge-transfer excitation of nitrobenzene in Figure~\ref{fig:nitrobenzene_convergence}, \gls{sgm} has been found to converge smoothly, but to a spurious solution in which the charge-transfer character is reduced by mixing of the hole with an initially occupied orbital~\cite{Carter2020}. Similar issues have recently been observed for long-range charge-transfer excitations in molecular dimers by Bogo and Stein~\cite{Bogo2024}.

Rather than attempting to compute \gls{oo} excited states as minima on the squared gradient landscape, a direct optimization of the energy can be carried out using saddle point search algorithms adapted from methods originally developed for locating transition states in rearrangements of atoms, corresponding to first order saddle points on potential energy surfaces. An efficient strategy is to employ quasi-Newton algorithms for unconstrained optimization that can develop negative Hessian eigenvalues. A limited-memory symmetric rank 1 (L-SR1) algorithm has been found to be particularly effective\cite{Ivanov2021, Levi2020b}. In early applications, direct optimization was combined with MOM to reduce the risk of variational collapse~\cite{Ivanov2021,Levi2020a,Levi2020b}. The generalized mode following (GMF) approach was later introduced, which does not use MOM~\cite{Schmerwitz2023}. GMF generalizes minimum mode following, commonly used in transition state searches, to saddle points of arbitrary order. To target a saddle point of order $n$, the eigenvectors ${\mathbf v}_i$ corresponding to the $n$ lowest eigenvalues of the electronic Hessian are determined using, e.g., numerical partial diagonalization, and the components of the gradient along these modes are inverted:
\begin{equation}
\mathbf g_{\rm mod}
=
\nabla_{\mathbf \kappa} E
-
2\sum_{i=1}^{n}
\mathbf v_i \mathbf v_i^\mr{T} \nabla_{\mathbf \kappa} E.
\end{equation}
Minimizing along \(\mathbf g_{\rm mod}\) corresponds to moving uphill along the \(n\) directions of negative curvature and downhill along all remaining directions, thereby converging to the target saddle point associated with the desired excited state. Figure ~\ref{fig:nitrobenzene_convergence} illustrate that both DO-MOM with L-SR1 and DO-GMF can converge a challenging charge-transfer state in nitrobenzene~\cite{Schmerwitz2023,Ivanov2021,Levi2020a,Levi2020b}. While DO-GMF involves a larger computational effort due to the need of computing the lowest eigenvectors of the electronic Hessian, it gives smoother and more systematic convergence. Importantly, the failure of DIIS-MOM and the success of DO-MOM for the charge-transfer excited states of nitrobenzene shows that MOM alone does not determine convergence. Rather, convergence is determined by the underlying orbital-optimization algorithm.

\begin{figure}
\centering
\includegraphics[width = 0.9\linewidth]{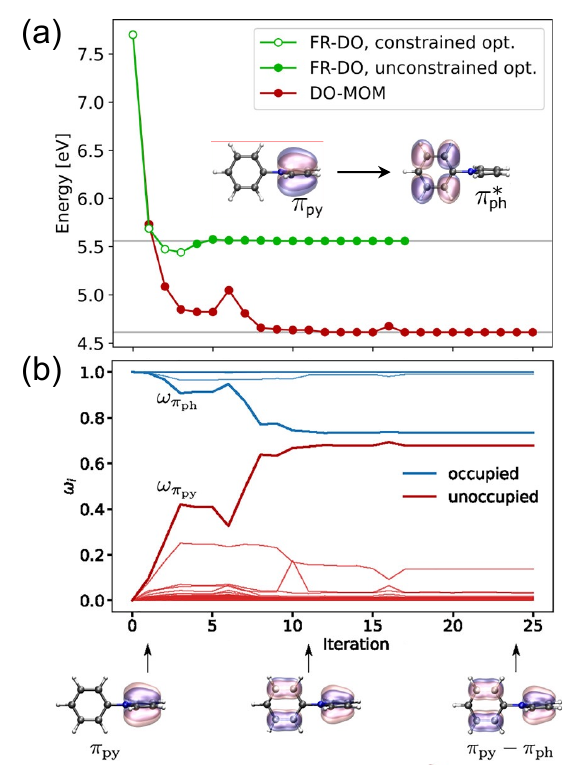}
\caption{(a) Convergence of the excitation energy for direct optimization (DO) calculations of an excited state of the \(N\)-phenylpyrrole molecule in a twisted geometry involving electron transfer from a $\pi$ orbital on the pyrrole group to a $\pi^*$ orbital on the phenyl ring. DO with maximum overlap method (DO-MOM) collapses to a low-energy solution with reduced charge-transfer character, while freeze-and-release DO (FR-DO) converges to the target charge-localized state. 
(b) Weights \(\omega_j\) used to assign the occupation numbers during the DO-MOM calculation (see eq. \ref{eq:mom}). During the optimization, an occupied \(\pi\) orbital initially localized on the phenyl ring and the unoccupied \(\pi\) hole orbital mix by approximately \(45^\circ\), leading to a charge-delocalized solution. Because the occupied and unoccupied orbitals do not exchange their maximum overlap ranking, MOM is unable to avoid this collapse.  
Adapted from Y. L. A. Schmerwitz, E. Selenius, and G. Levi, ``Freeze-and-Release Direct Optimization Method for Variational Calculations of Excited Electronic States'', J. Chem. Theory Comput. \textbf{22}, 3571--3584 (2026). Published by American Chemical Society under the Creative Commons Attribution 4.0 International License (CC BY 4.0).}
\label{fig:twisted_pp_mom_failure}
\end{figure}

The L-SR1 and GMF direct optimization methods critically depend on identifying the degrees of freedom along which the energy should be maximized. Figure \ref{fig:initial_guess_spo} shows that the common initial guess of ground-state orbitals with nonaufbau occupations does not always provide a reliable estimate of the saddle point order of the final solution. Figure \ref{fig:twisted_pp_mom_failure} illustrates this issue for a charge-transfer excitation in the \(N\)-phenylpyrrole molecule involving electron transfer from the pyrrole to the benzene ring\cite{Schmerwitz2026}. 
There, a DO-MOM calculation with L-SR1 started from ground-state orbitals with changed occupations collapses to a solution with reduced charge-transfer character because the hole orbital, localized on the pyrrole, mixes with an occupied orbital localized on the phenyl. The problem is that the energy should be maximized along the rotation mixing these two orbitals, while an estimate based on a diagonal approximation of the Hessian predicts a positive curvature\cite{Schmerwitz2026}. Once again, MOM alone is not able to prevent this failure. In this case, the orbitals mix by $\sim45^\circ$ and since occupied and unoccupied orbitals do not exchange their maximum overlap ranking with respect to the initial guess (see Figure \ref{fig:twisted_pp_mom_failure}), MOM does not detect a collapse. The freeze-and-release direct optimization (FR-DO)~\cite{Schmerwitz2026} addresses this problem by performing a first, constrained optimization step, where the hole and particle orbitals defining the target excitation are kept frozen while all other orbitals are relaxed. This step prevents collapse and leads to a more reliable estimate of the directions of negative curvature associated with the target saddle point (see Figure \ref{fig:initial_guess_spo}). In a second step, the constraints are removed and an unconstrained saddle point optimization is performed. This two-step procedure does not use MOM and, as shown in Figure~\ref{fig:twisted_pp_mom_failure}, converges the challenging charge-transfer excited state of \(N\)-phenylpyrrole, avoiding the spurious charge-delocalized state found by DO-MOM. A first step of constrained optimization is also needed for a reliable estimate of the saddle point order to target in a DO-GMF calculation. 

The freeze-and-release idea has a precedent in an earlier SCF strategy for \gls{oo} Hartree-Fock calculations of excited states~\cite{Obermeyer2021}, where it was employed together with DIIS and MOM, and is currently being extended in several directions. Stein and co-workers combined freeze-and-release with \gls{sgm} and applied the approach to long-range intermolecular charge-transfer excitations in large donor--acceptor systems~\cite{Bogo2025}. Qin and Suo recently proposed a freeze-and-release scheme for core excitations, in which the frozen orbitals are transition orbitals obtained from a preliminary TDA calculation rather than canonical ground-state orbitals~\cite{Qin2026-mf}. This was shown to improve convergence over MOM-based calculations for core excited states characterized by strong orbital mixing. 

A final example illustrating the advantages of methods specifically designed to locate saddle points is provided by the lowest doubly excited state of H$_2$, discussed in the previous section (see Figure \ref{fig:h2_curves_sp_order}). The higher-energy curve with the correct long-range behavior is given by a solution that is a second-order saddle point at all geometries and breaks the spatial symmetry at long bond lengths. DO-GMF can follow the desired solution by targeting a second-order saddle point, and thereby provides a systematic way to compute the energy curve along the bond-stretching coordinate~\cite{Schmerwitz2024,Schmerwitz2023}. By contrast, MOM alone cannot drive the calculation toward the symmetry-broken solution after the symmetry breaking onset, because reaching this solution requires mixing of the occupied and unoccupied orbitals rather than a reassignment of occupations. This again shows that MOM is best viewed as a useful  device to track the occupation numbers and monitor variational collapse, not as a general solution to the excited-state optimization problem.

Finally, several constraint-based \gls{oo} density functional approaches have recently emerged, in which excited states are optimized under additional constraints~\cite{Lemke2026,Pham2025,Kussmann2024,Lemke2024-tg,Stella2022,Roychoudhury2020,Ramos2018,Evangelista2013,Gavnholt2008}. The constraints stabilize the excited state optimization, preventing variational collapse by construction. While promising, these methods are not reviewed here, as the focus is on fully variational methods that seek solutions of the unconstrained electronic energy landscape. 

\subsection{Methods for open-shell singlet excited states \label{sec:methods_roks}}
Open-shell singlet excited states represent an important and vast class of electronic excitations, including all singly excited states of molecules with a singlet closed-shell ground state. The wave function of an open-shell singlet state is intrinsically multi-determinantal. Here, we discuss the case where there are two open-shell orbitals, \(a\) and \(b\), with the corresponding spin-adapted \gls{csf} being
\begin{equation}\label{eq:singlet_csf}
    \ket{\Psi_\mathrm{S}} =
    \frac{1}{\sqrt{2}}
    \left(
    \ket{a\bar{b}} +
    \ket{b\bar{a}}
    \right),
\end{equation}
where the bar indicates a beta spin orbital, and the common set of doubly occupied ``core'' orbitals has been omitted from the Slater determinants for simplicity. The spin-summed one-particle reduced density matrix, $D_{pq}$, in the molecular orbital basis associated with this \gls{csf} is diagonal, with occupation numbers two for the doubly occupied core orbitals and one for the two open-shell orbitals, with all off-diagonal blocks equal to zero,
\begin{equation}
    D_{ij} = 2\delta_{ij}, \qquad
    D_{ab} = \delta_{ab}, \qquad
    D_{uv} = 0.
    \label{eq:open_shell_singlet_density_matrix_blocks}
\end{equation}
The corresponding real-space electron density is therefore
\begin{align}
    n_\mathrm{S}(\mathbf r)
    &=
    \sum_{pq}
    D_{pq}
    \psi_p^*(\mathbf r)\psi_q(\mathbf r) \nonumber \\
    &=
    2 \sum_{i\in{\rm core}}
    \left|\psi_i(\mathbf r)\right|^2
    +
    \left|\psi_a(\mathbf r)\right|^2
    +
    \left|\psi_b(\mathbf r)\right|^2 .
    \label{eq:open_shell_singlet_density}
\end{align}
Open-shell singlet states require special treatment within \gls{oo} density functional calculations, because the conventional KS approach is formulated in terms of a single determinant. 

Here, we attempt to provide an overview of the most common \gls{oo} density functional approaches to deal with open-shell singlet excited states with two open-shell orbitals. This is not intended as a fully comprehensive review of all \gls{oo} methods for open-shell singlets, which constitute a rapidly expanding area, with new approaches continuing to emerge\cite{Trushin2026Chemrxiv}. It is also important to emphasize that not all excited states require such specialized treatments. For example, triplet excited states and certain doubly excited states reached from closed-shell singlet ground states can be described within a standard KS framework, as can certain singly excited states of systems with open-shell singlet ground states.

\subsubsection{Spin purification} 
\begin{figure}
\centering
\includegraphics[width = 1\linewidth]{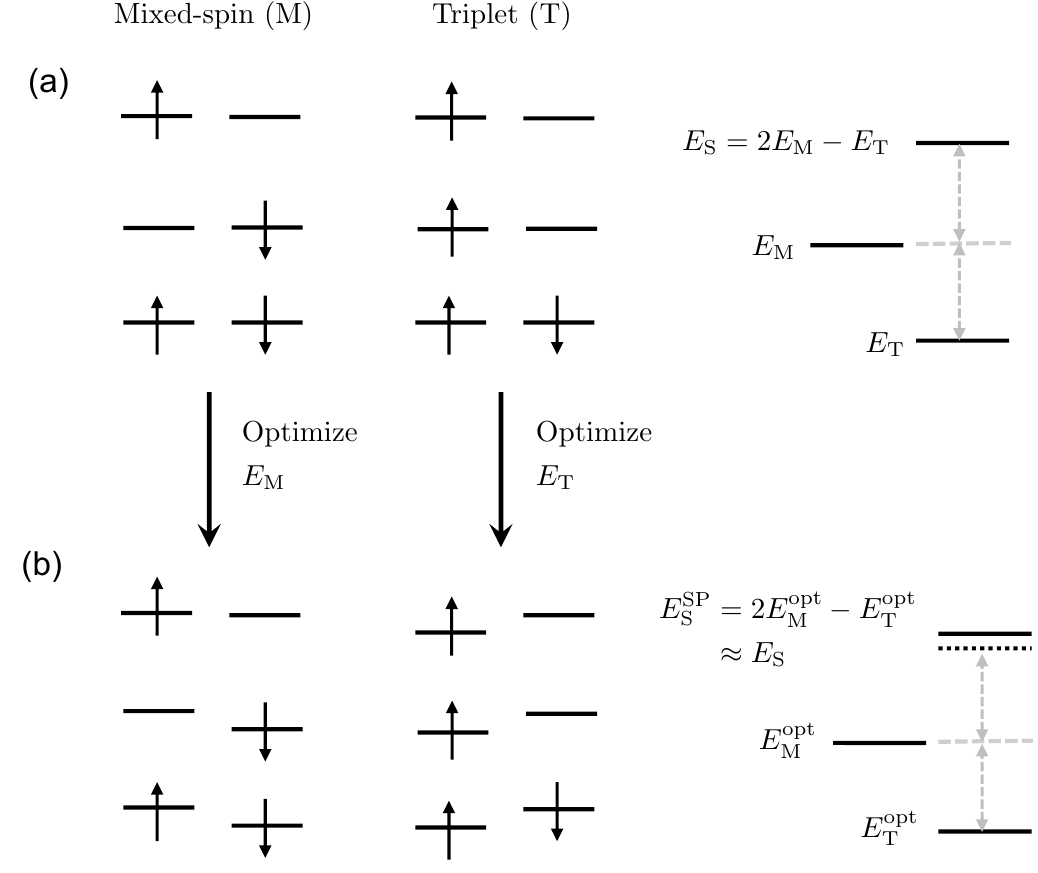}
\caption{
Schematic representation of spin-unrestricted orbital-optimized (\gls{oo}) calculations with spin purification for open-shell singlet excited states. (a) For a fixed common set of orbitals, the spin-mixed (M) determinant is an equal mixture of the singlet and \(M_S=\pm 0\) triplet (T) wave functions. (b) In practical \gls{oo} calculations, the mixed-spin and triplet \(M_S=\pm 1\) determinants are optimized separately, and spin purification is approximate. In the restricted open-shell Kohn-Sham approaches discussed in section \ref{sec:roks}, a single set of restricted orbitals is optimized.
}
\label{fig:spin_purification}
\end{figure}

One of the earliest and still most used approaches for estimating the energy and properties of open-shell singlet states is the spin-purification method, as illustrated in Figure \ref{fig:spin_purification}. This approach starts from the observation that the single determinant \(\ket{a\bar{b}}\) with two unpaired electrons is spin-mixed, as it is not an eigenfunction of the spin squared operator, \(\hat{S}^2\), and it has an expectation value \(\langle \hat{S}^2 \rangle_\mathrm{M} = 1\), corresponding to an equal mixture of singlet and \(M_S=0\) triplet wave functions,
\begin{equation}
    \ket{a\bar{b}}
    =
    \frac{1}{\sqrt{2}}
    \left(
    \ket{\Psi_\mathrm{S}}
    +
    \ket{\Psi_{\mathrm{T},0}}
    \right),
    \label{eq:spin_mixed_det}
\end{equation}
where
\begin{equation}
    \ket{\Psi_{\mathrm{T},0}}
    =
    \frac{1}{\sqrt{2}}
    \left(
    \ket{a\bar{b}} -
    \ket{b\bar{a}}
    \right).
    \label{eq:triplet_ms0_csf}
\end{equation}
In practice, \(\langle \hat{S}^2 \rangle_\mathrm{M} \approx 1\) if the spin-mixed state $\ket{a\bar{b}}$ is optimized in a self-consistent calculation, since the optimized orbitals will be different from the orbitals of the singlet and triplet wave functions.

For any spin-independent operator \(\hat{O}\), the expectation value with the spin-mixed determinant is
\begin{align}
    \langle \hat{O} \rangle_\mathrm{M}
    &=
    \langle a\bar{b} | \hat{O} | a\bar{b} \rangle \nonumber \\
    &=
    \frac{1}{2}
    \left(
    \langle \Psi_\mathrm{S} | \hat{O} | \Psi_\mathrm{S} \rangle
    +
    \langle \Psi_{\mathrm{T},0} | \hat{O} | \Psi_{\mathrm{T},0} \rangle
    \right. \nonumber \\
    &\qquad\left.
    +
    \langle \Psi_\mathrm{S} | \hat{O} | \Psi_{\mathrm{T},0} \rangle
    +
    \langle \Psi_{\mathrm{T},0} | \hat{O} | \Psi_\mathrm{S} \rangle
    \right) \nonumber \\
    &= \frac{1}{2} 
        \left(
    \langle \Psi_\mathrm{S} | \hat{O} | \Psi_\mathrm{S} \rangle
    +
    \langle \Psi_{\mathrm{T},0} | \hat{O} | \Psi_{\mathrm{T},0} \rangle
    \right)
    \nonumber \\
    &= \frac{1}{2} 
    \left( \langle \hat{O} \rangle_\mathrm{S} + \langle \hat{O} \rangle_\mathrm{T} \right) ,
    \label{eq:spin_purification_operator}
\end{align}
and therefore
\begin{equation}
    \langle \hat{O} \rangle_\mathrm{S}
    =
    2\langle \hat{O} \rangle_\mathrm{M}
    -
    \langle \hat{O} \rangle_\mathrm{T}.
    \label{eq:spin_purification_general}
\end{equation}
Applied to the energy, this gives the commonly used spin-purification formula of Ziegler, Rauk, and Baerends~\cite{Ziegler1977},
\begin{equation}
    E_\mathrm{S}
    =
    2E_\mathrm{M} - E_\mathrm{T}.
    \label{eq:spin_purification_energy}
\end{equation}
Similarly, for the permanent electric dipole moment,
\begin{equation}
    \boldsymbol{\mu}_\mathrm{S}
    =
    2\boldsymbol{\mu}_\mathrm{M} - \boldsymbol{\mu}_\mathrm{T}.
    \label{eq:spin_purification_dipole}
\end{equation}

The same approach can be used for transition properties. For the \gls{tdm} between the ground state, $\ket{\Psi_0}$, and the open-shell singlet excited state, since the \gls{tdm} between a singlet state and the \(M_S=0\) triplet vanishes, one obtains that
\begin{equation}
    \langle \Psi_\mathrm{S} | \hat{\boldsymbol{\mu}} | \Psi_0 \rangle
    =
    \sqrt{2}
    \langle a\bar{b} | \hat{\boldsymbol{\mu}} | \Psi_0 \rangle ,
    \label{eq:spin_purification_tdm}
\end{equation}
where $\hat{\boldsymbol{\mu}}$ is the electric dipole operator (see eq. \ref{eq:elect_dipole}).

In practical \gls{oo} density functional calculations, two separate spin-unrestricted calculations are performed to apply spin purification, one for the spin-mixed determinant, \(\ket{a\bar{b}}\) or \(\ket{b\bar{a}}\), and one for the triplet determinant with \(M_S=\pm 1\), \(\ket{ab}\) or \(\ket{\bar{a}\bar{b}}\). The latter is used to compute $\langle \hat{O} \rangle_\mathrm{T}$ instead of using the \(M_S=\pm 0\) state, which is justified when the Hamiltonian is spin independent. Spin purification in \gls{oo} density functional calculations is an approximate procedure. The derivation assumes that the spin-mixed determinant is an exact equal mixture of the open-shell singlet and triplet \(M_S=0\) states, and that the singlet, spin-mixed, and triplet quantities are evaluated with the same set of orbitals. These conditions are not generally satisfied in \gls{oo} calculations, where the spin-mixed and high-spin triplet states are optimized independently. The method also doubles the number of calculations required for a single-point energy computation as well as excited-state geometry optimizations and molecular dynamics simulations, which require spin purification of the atomic forces. For this reason, geometry optimizations and dynamics are sometimes performed directly on the spin-mixed surface\cite{Pradhan2018}, or on the triplet surface\cite{Dohn2014}, relying on the observation that the singlet and triplet potential energy surfaces are often approximately parallel\cite{Duchstein2012}.

\subsubsection{Restricted open-shell Kohn-Sham approaches}\label{sec:roks}
In recent years, a host of \gls{oo} density functional approaches have emerged to compute open-shell singlet excited states in a single variational calculation, rather than from separate spin-mixed and triplet calculations as in post-SCF spin purification. These methods differ primarily in how the xc contribution of the open-shell singlet is approximated. Although not all of them are referred to as \gls*{roks} in the original literature, we choose to use this terminology here for different methods, to emphasize their common structure, namely a KS-like orbital-optimized description based on restricted orbitals with open-shell occupations.

A useful starting point for understanding the different \gls{roks} formulations currently in use are the Hartree--Fock expressions of the energy for the open-shell singlet, triplet and spin-mixed wave functions:
\begin{align}
    E_\mathrm{S}^\mathrm{HF}
    &=
    T + V_\mathrm{ext}[n] + E_\mathrm{H}[n]
    + E_\mathrm{x}
    + K_{ab},
    \nonumber \\
    E_\mathrm{T}^\mathrm{HF}
    &=
    T + V_\mathrm{ext}[n] + E_\mathrm{H}[n]
    + E_\mathrm{x}
    - K_{ab},
    \nonumber \\
    E_\mathrm{M}^\mathrm{HF}
    &=
    T + V_\mathrm{ext}[n] + E_\mathrm{H}[n]
    + E_\mathrm{x} .
    \label{eq:HF_energies_open_shell}
\end{align}
When the states are not optimized independently, the spin-summed one-particle density matrix, and therefore the total density \(n(\mathbf r)\), is the same for the open-shell singlet, the triplet, and the spin-mixed determinant. The kinetic, external potential, Hartree, and $E_\mathrm{x}$ Fock exchange contributions are therefore common to all three states. The kinetic energy is given by
\begin{equation}
    T
    =
    \sum_{pq} D_{pq} t_{pq},
    \qquad
    t_{pq}
    =
    -\frac{1}{2}
    \int
    \psi_p^*(\mathbf r)
    \nabla^2
    \psi_q(\mathbf r)
    d\mathbf r .
    \label{eq:kinetic_energy_density_matrix}
\end{equation}
The expression of the Hartree energy is the same as eq. \ref{eq:Hartree}, while \(E_\mathrm{x}\) denotes the Fock exchange contributions arising from the diagonal matrix elements of the electron-electron interaction operator, i.e.
\(\langle a\bar b|\hat{V}_{ee}|a\bar b\rangle\) and
\(\langle b\bar a|\hat{V}_{ee}|b\bar a\rangle\), after the Hartree contribution has been separated out. For two open-shell orbitals, $a$ and $b$, \(E_\mathrm{x}\) is given by 
\begin{equation}
    E_\mathrm{x}
    =
    -
    \sum_{i\in{\rm core}} K_{ij}
    -
    \sum_{i\in{\rm core}}
    \left(
    K_{ia}
    +
    K_{ib}
    \right)
    - 
    \frac{1}{2} \left(K_{aa}+K_{bb}\right),
    \label{eq:diagonal_exchange}
\end{equation}
where
\begin{equation}
    K_{pq}
    =
    \iint
    \frac{
    \psi_p^*(\mathbf r)
    \psi_q^*(\mathbf r')
    \psi_q(\mathbf r)
    \psi_p(\mathbf r')
    }
    {|\mathbf r-\mathbf r'|}
    d\mathbf r d\mathbf r' .
    \label{eq:exchange_integral}
\end{equation}
The additional term \(K_{ab}\) in the expressions for the singlet and triplet energy in eq. \ref{eq:HF_energies_open_shell} is the exchange coupling that originates from the off-diagonal matrix elements of the electron--electron interaction operator between the two determinants entering the CSF, i.e.
\(\langle a\bar b|\hat{V}_{ee}|b\bar a\rangle\). This term is absent from the energy expression for the spin-mixed determinant, appears with a negative sign in the triplet energy, and appears with a positive sign in the open-shell singlet energy. Consequently,
\begin{equation}
    E_\mathrm{M}^\mathrm{HF}
    =
    \frac{1}{2}
    \left(
    E_\mathrm{S}^\mathrm{HF}
    +
    E_\mathrm{T}^\mathrm{HF}
    \right),
    \label{eq:hf_mixed_average}
\end{equation}
or equivalently
\begin{equation}
    E_\mathrm{S}^\mathrm{HF}
    =
    2E_\mathrm{M}^\mathrm{HF}
    -
    E_\mathrm{T}^\mathrm{HF}.
    \label{eq:hf_spin_purification}
\end{equation}
Thus, the usual spin-purification formula is obtained from the Hartree--Fock energy expressions for a single set of orbitals. The singlet--triplet gap is equal to \(2K_{ab}\), while the spin-mixed determinant lies halfway between the singlet and triplet states.

\begin{table*}[t]
    \centering
    \caption{
    Overview of restricted open-shell Kohn-Sham (ROKS) approaches for open-shell singlet excited states with two open-shell orbitals, $a$ and $b$, described by the configureation state function of eq. \ref{eq:singlet_csf}. The orbital occupations in the spin-mixed (M) and triplet (T) single Slater determinants are shown in Figure \ref{fig:spin_purification}. The second column indicates how the exchange and correlation contribution to the energy is described. The last column gives the singlet--triplet splitting when orbital relaxation effects between the singlet and triplet states are neglected. In all ROKS methods, a single set of spin-restricted orbitals is optimized. The Hartree-Fock expressions for the open-shell singlet state, where exchange is treated exactly, are also included. Originally, the term ROKS was used to refer to the method designated here as SP-ROKS, but it is used here to refer to other approaches for open-shell singlet states based on restricted orbitals as well, which also reflects a recent trend in the literature.
    }
    \label{tab:roks_overview}
    \begin{tabular}{lll}
        \hline
        Method
        &
        \begin{tabular}{@{}l@{}}
        Exchange--correlation treatment
        \end{tabular}
        &
        Singlet-triplet splitting (approx.)
        \\
        \hline

        Hartree-Fock
        &
        \(E_\mathrm{x}+K_{ab}\)
        &
        \(2K_{ab}\)
        \\[6pt]

        SP-ROKS\textsuperscript{a}
        &
        \(2E_\mathrm{xc}[n_\mathrm{M}^{\alpha},n_\mathrm{M}^{\beta}]
        -
        E_\mathrm{xc}[n_\mathrm{T}^{\alpha},n_\mathrm{T}^{\beta}]\)
        &
        \(2\left(
        E_\mathrm{xc}[n_\mathrm{M}^{\alpha},n_\mathrm{M}^{\beta}]
        -
        E_\mathrm{xc}[n_\mathrm{T}^{\alpha},n_\mathrm{T}^{\beta}]
        \right)\)
        \\[6pt]

        SU-ROKS\textsuperscript{b}
        &
        \(E_\mathrm{xc}[n]\)
        &
        \(E_\mathrm{xc}[n]
        -
        E_\mathrm{xc}[n_\mathrm{T}^{\alpha},n_\mathrm{T}^{\beta}]\)
        \\[6pt]

        ESMF-ROKS\textsuperscript{c}
        &
        \(E_\mathrm{xc}[n]+K_{ab}\)
        &
        \(E_\mathrm{xc}[n]
        -
        E_\mathrm{xc}[n_\mathrm{T}^{\alpha},n_\mathrm{T}^{\beta}]
        +
        K_{ab}\)
        \\[6pt]

        EDFT-ROKS\textsuperscript{d}
        &
        \(E_\mathrm{xc}[n_\mathrm{T}^{\alpha},n_\mathrm{T}^{\beta}]+2K_{ab}\)
        &
        \(2K_{ab}\)
        \\
        \hline
    \end{tabular}

    \vspace{0.5em}
    \begin{flushleft}
    \footnotesize
    \textsuperscript{a}SP: spin-purification.\cite{Frank1998, Filatov1998}
    \textsuperscript{b}SU: spin-unpolarized.\cite{Birgisson2025, Sinyavskiy2025, Vandaele2022b, Kumar2022, Levi2020pccp, Malis2020, Levi2018, Himmetoglu2012, Maurer2011}
    \textsuperscript{c}ESMF: excited-state mean-field.\cite{Zhao2019}
    \textsuperscript{d}EDFT: ensemble density functional theory.\cite{Gould2026}
    \end{flushleft}
\end{table*}
Within Hartree-Fock, the energy expression of the open-shell singlet wave function is exact. The difficulty in moving from Hartree--Fock to a KS approach is that approximate exchange--correlation functionals are not designed to capture exchange and correlation for the open-shell singlet CSF. In the classical ROKS approach developed independently by Parrinello and co-workers~\cite{Frank1998} and by Filatov and Shaik~\cite{Filatov1998}, the singlet energy is written as
\begin{equation}\label{eq:roks_classic}
    E_\mathrm{S}^\mathrm{SP-ROKS}
    =
    2E_\mathrm{M}^\mathrm{KS}[\{\psi_p\}]
    -
    E_\mathrm{T}^\mathrm{KS}[\{\psi_p\}],
\end{equation}
where the energy of the spin-mixed and triplet states is evaluated using the same restricted set of spatial orbitals. The term ROKS has been originally used, and is still most commonly used, to refer to this method. Here, we choose to label it as SP-ROKS, where SP indicates that the approach rests on the spin-purification formula, in order to distinguish it from other methods also designated as ROKS. 

In SP-ROKS, the orbitals are optimized using the expression in eq. \ref{eq:roks_classic} as the objective function. Because the spin-mixed and triplet determinants have the same reduced density matrix, the SP-ROKS singlet energy can be written as
\begin{equation}
    E_\mathrm{S}^\mathrm{SP-ROKS}
    =
    T
    +
    V_\mathrm{ext}[n]
    +
    E_\mathrm{H}[n]
    +
    2E_\mathrm{xc}
    \left[
    n_\mathrm{M}^\alpha,
    n_\mathrm{M}^\beta
    \right]
    -
    E_\mathrm{xc}
    \left[
    n_\mathrm{T}^\alpha,
    n_\mathrm{T}^\beta
    \right] ,
    \label{eq:classical_roks_energy}
\end{equation}
Comparing this expression with eq. \ref{eq:HF_energies_open_shell}, the classical ROKS approach can be seen as replacing $E_x + K_{ab}$ in the Hartree-Fock energy expression with a linear combination of approximate exchange--correlation functionals evaluated for the spin-mixed and triplet spin densities. Neglecting orbital relaxation effects, the singlet-triplet gap is given by $2\left(E_\mathrm{xc} \left[ n_\mathrm{M}^\alpha, n_\mathrm{M}^\beta\right] - E_\mathrm{xc}\left[n_\mathrm{T}^\alpha,n_\mathrm{T}^\beta\right]\right)$, which is consistent with the spin-mixed state being approximately half way between the singlet and triplet. 

The classical ROKS energy expression is not invariant under unitary rotation between unequally occupied orbitals. As a result, the corresponding one-electron equations have different effective operators for the closed-shell and open-shell orbitals\cite{Frank1998, Filatov1998}. In diagonalization-based SCF implementations, Roothaan's coupling-operator technique can be used to construct a generalized effective Fock matrix~\cite{Kowalczyk2013, Filatov1998, Roothaan1960}. Alternatively, the ROKS functional can be optimized directly with respect to orbital rotations that preserve their orthonormality~\cite{Frank1998} (see section \ref{sec:oo_algorithms}), in a manner similar to other orbital-density-dependent functionals, such as self-interaction-corrected functionals~\cite{Ivanov2021, Perdew1981}. Atomic forces can be obtained from a single ROKS calculation~\cite{Kowalczyk2013}, rather than from separate spin-mixed and triplet calculations as in post-SCF spin purification.

Several recent works, in particular by Luber and co-workers, use a simpler ROKS formulation\cite{Birgisson2025, Sinyavskiy2025, Vandaele2022b, Kumar2022, Levi2020pccp, Malis2020, Levi2018, Himmetoglu2012, Maurer2011}, where the xc energy is evaluated directly from the spin-unpolarized (SU) density of the open-shell singlet,
\begin{equation}
    E_\mathrm{S}^{\mathrm{SU-ROKS}}
    =
    T
    +
    V_\mathrm{ext}[n]
    +
    E_\mathrm{H}[n]
    +
    E_\mathrm{xc}[n],
    \label{eq:density_based_roks_energy}
\end{equation}
where \(E_\mathrm{xc}[n] = E_\mathrm{xc}[n/2,n/2]\). In terms of the Hartree--Fock decomposition discussed above, this approximation may be viewed as replacing the full open-shell exchange contribution \(E_\mathrm{x}^{\mathrm{diag}}+K_{ab}\) by the approximate xc energy evaluated on the total open-shell singlet density. This amounts to assuming that the energy is a functional of the spin-summed one-particle reduced density matrix, thereby neglecting an explicit coupling between the two determinant of the singlet open-shell CSF~\cite{Cernatic2022}. 

This second type of restricted ROKS formulation is simpler to implement than the spin-purified ROKS expression of eq.~\eqref{eq:classical_roks_energy}, because it requires only a conventional spin-restricted KS energy evaluation with occupation numbers of 2 for the core orbitals and 1 for the open-shell orbitals. Despite its simplicity, the method still requires care in the orbital optimization, as the energy is still not invariant under rotations between orbitals with different occupation numbers. Once a variational solution is obtained, computing atomic forces is straightforward within this approach. As a result, in recent years the SU-ROKS approach has been widely used for excited-state geometry optimizations and molecular dynamics simulations, where repeated evaluations of energy and atomic forces are required\cite{Birgisson2025, Vandaele2022b, Levi2020pccp, Malis2020, Levi2018, Maurer2011}.

A third, ROKS-like construction has been proposed by Zhao and Neuscamman\cite{Zhao2019} in the context of a density functional extension to excited-state mean-field theory (ESMF)\cite{Shea2018}. In this approach, the energy is written as
\begin{equation}
    E_\mathrm{S}^{\mathrm{ESMF\text{-}ROKS}}
    =
    T
    +
    V_\mathrm{ext}[n]
    +
    E_\mathrm{H}[n]
    +
    E_\mathrm{xc}[n]
    +
    K_{ab}.
    \label{eq:esmf_roks_energy}
\end{equation}
Thus, this approximation can be seen as replacing the open-shell exchange contribution in the Hartree-Fock expression with the xc energy evaluated with the spin-unpolarized density functional plus the off-diagonal exchange coupling \(K_{ab}\) between the two determinants of the open-shell CSF, here included explicitly. Initial applications showed good performance for long-range charge-transfer excitations after orbital relaxation (see also \ref{sec:ct_states}). However, for such excitations, \(K_{ab}\) tends to vanish as the distance between hole and electron orbitals is large, so the method reduces to the SU-ROKS approach. For valence and Rydberg excited states, ESMF-ROKS has been found to overestimate the excitation energy with common local and hybrid functionals relative to coupled-cluster references~\cite{Zhao2019}.

A fourth ROKS-like approach has been proposed by Gould, Kronik, and Pittalis in the context of ensemble DFT (EDFT)~\cite{Gould2026}. For the pure density functionals, the open-shell singlet energy is written as
\begin{equation}
    E_\mathrm{S}^{\mathrm{EDFT\text{-}ROKS}}
    =
    T
    +
    V_\mathrm{ext}[n]
    +
    E_\mathrm{H}[n]
    +
    E_\mathrm{xc}
    \!\left[
    n_\mathrm{T}^{\alpha},
    n_\mathrm{T}^{\beta}
    \right]
    +
    2K_{ab}.
    \label{eq:edft_roks_energy}
\end{equation}
Thus, this formulation evaluates the xc energy using the spin-polarized density of the corresponding triplet configuration, for which well-developed approximations are available, and adds twice the off-diagonal exchange coupling \(K_{ab}\). If orbital relaxation effects are neglected, the singlet--triplet splitting corresponds to \(2K_{ab}\), as in Hartree-Fock theory.

The different ROKS-like formulations discussed in this section are summarized in Table~\ref{tab:roks_overview}. We have focused here on the basic structure of these approximations when pure density functionals are used, and have not discussed the additional complications that arise with hybrid functionals, which include a fraction of exact Fock exchange. For a discussion of these issues see, e.g., references ~\citen{Malis2026} and ~\citen{Gould2026}.

Finally, we should note that a recent potential-averaged (pa) KS approach for open-shell singlets has been proposed by Trushin, Bertleff, and G\"orling~\cite{Trushin2026Chemrxiv}. This method computes the energy using the spin-purification formula of eq.~\ref{eq:roks_classic}, but determines the orbitals from a canonical KS equation with an averaged xc potential
\begin{equation}
    v_\mathrm{xc}^{\mathrm{pa}}
    =
    \frac{1}{2}
    \left(
    2v_\mathrm{xc}^{\mathrm{M},\alpha}
    +
    2v_\mathrm{xc}^{\mathrm{M},\beta}
    -
    v_\mathrm{xc}^{\mathrm{T},\alpha}
    -
    v_\mathrm{xc}^{\mathrm{T},\beta}
    \right).
    \label{eq:pa_oss_vxc}
\end{equation}
This construction avoids the complications that arise in classical ROKS formulations from the lack of invariance of the energy under rotations between unequally occupied orbitals. However, the evaluation of analytical forces is less straightforward, because the final spin-purified energy is not stationary with respect to the orbital rotations.

\subsection{Calculation of absorption and emission spectra \label{sec:calculation_spectra}}
Any electronic structure method designed to describe excited states is ultimately expected to provide access to spectroscopic observables. In practice, this requires evaluating transition properties that provide a direct link between the theoretical excited-state description and measurable signals. In the case of absorption and emission spectra, essentially all \gls*{oo} density functional approaches evaluate intensities in a wave-function-like manner by using the \gls{ks} wave function as an approximation to the exact wave function. This procedure is not formally rigorous, because the \gls{ks} system is an auxiliary object. Nevertheless, several studies~\cite{Yang2026-et, Sinyavskiy2025,Toffoli2022,Bourne2021,Hait2020radicals} have shown that reasonable transition properties can be obtained in this way.

Within the electric-dipole approximation, the intensity of an electronic transition is commonly expressed in terms of the oscillator strength. In the length gauge, the oscillator strength for a transition from an initial state $k$ to a final state $k^\prime$ is given, in atomic units, by
\begin{equation}
f^{kk^\prime} = \frac{2}{3} \Delta E ^{kk^\prime}  \left|\bm{\mu}^{kk^\prime}\right|^2 ,
\end{equation}
where $\Delta E ^{kk^\prime}=E ^{k^\prime}-E^k$ is the excitation energy and $\bm{\mu}^{kk^\prime}$ is the \gls{tdm}. The latter is defined as
\begin{equation}\label{eq:tdm}
\bm{\mu}^{kk^\prime}
=
\left\langle \Psi ^k \middle| \hat{\bm{\mu}} \middle| \Psi ^{k^\prime} \right\rangle .
\end{equation}
For a molecular system with $N$ electrons and $M$ nuclei, the electric dipole moment operator is
\begin{align}\label{eq:elect_dipole}
\hat{\bm{\mu}} =
-e \sum_i^{N} \bm{r}_i + e \sum_a^{M} Z_{a} \bm{R}_a \, ,
\end{align}
where $\bm{r}_i$ denotes the position of electron $i$, while $Z_a$ and $\bm{R}_a$ are the charge and position of nucleus $a$, respectively.

In \gls*{oo} density functional methods, $\bm{\mu}^{kk^\prime}$ is evaluated using the ground- and excited-state KS wave functions optimized in separate calculations. Since ground- and excited-state optimized orbitals are generally mutually nonorthogonal, this involves an additional complication that is absent when all states are represented in a common orthonormal orbital basis, as transition matrix elements cannot be evaluated directly using the usual Slater--Condon rules. The nonorthogonality problem, and the corresponding approaches for evaluating transition dipole moments, are discussed in the following section.

For open-shell singlet excited states, when a \gls{roks} formulation is employed, 
the \gls{tdm} can be evaluated using the open-shell CSF of eq. \ref{eq:singlet_csf}. In the case of spin-unrestricted calculations, instead, the excitation energy entering the oscillator strength is typically spin purified using eq.~\ref{eq:spin_purification_energy}, which requires two calculations, one for the spin-mixed determinant and one for the corresponding triplet determinant. In this case, as shown in eq. \ref{eq:spin_purification_tdm}, the \gls{tdm} itself is evaluated from the spin-mixed solution and does not require the triplet explicitly. 

\subsubsection{Transition dipole moment \label{sec:tdm}}
The calculation of \glspl{tdm} between nonorthogonal states obtained in \gls{oo} calculations can be carried out using L{\"o}wdin's rules for matrix elements of nonorthogonal determinants~\cite{Figari1985-ds,Lowdin1955-ve}. The matrix element of a one-particle operator $\hat O$ between two Slater determinants is expressed as
\begin{equation}\label{eq:Lowdin_NOCI}
\langle \Phi ^k | \hat O | \Phi ^{k^{\prime}}\rangle
=
\sum _{ij} O _{ij} ^{k k^{\prime}} \operatorname{cof}(\bm S^{k k^{\prime}})_{ij} ,
\end{equation}
where $\ket{\Phi ^k}$ and $\ket{\Phi ^{k^{\prime}}}$ are Slater determinants constructed from two sets of occupied spin-orbitals, $\{\psi _i ^k\}$ and $\{\psi _j ^{k^{\prime}}\}$, respectively. The one-particle matrix elements, $O _{ij}^{k k^{\prime}}$, are defined as
\begin{equation}
O _{ij}^{k k^{\prime}}
=
\expv{\psi _i ^k}{\hat O}{\psi _j ^{k^{\prime}}} \, ,
\end{equation}
and $\operatorname{cof}(\bm S^{k k^{\prime}})$ denotes the cofactor matrix associated with the matrix $\bm S^{k k^{\prime}}$ of overlaps between the molecular orbitals, $S _{ij}^{k k^{\prime}}=\bk{\psi _i ^k}{\psi _j ^{k^\prime}}$. The cofactor matrix is equal to the transpose of the adjugate matrix, 
\begin{equation}
\operatorname{cof}(\bm S^{kk^\prime})_{ij}
=
\operatorname{adj}(\bm S^{kk^\prime})_{ji}
=
\det(\bm S^{kk^\prime})
\left(\bm S^{kk^\prime}\right)^{-1}_{ji},
\end{equation}
with $\det (\bm S^{k k^{\prime}})$ corresponding to the overlap $\bk{\Phi ^k}{\Phi ^{k^\prime}}$ between the two Slater determinants. In some works~\cite{Yang2026-et,Toffoli2022,Bourne2021}, eq.~\eqref{eq:Lowdin_NOCI} is written in terms of the adjugate of the overlap matrix. However, as also pointed out by Lemke et al.~\cite{Lemke2024-tg}, the transpose of the adjugate, i.e. the cofactor matrix, should be used instead~\cite{Figari1985-ds,Lowdin1955-ve}. This distinction is important because the overlap matrix $\bm S^{kk^\prime}$ is not generally symmetric.

For unrestricted calculations, the overlap matrix is block diagonal:
\begin{equation}
\bm S^{kk'}=
\begin{pmatrix}
\bm S^{kk'}_{\alpha } & 0\\
0 & \bm S^{kk'}_{\beta }
\end{pmatrix} \,.
\end{equation}
Since the determinant of a block-diagonal matrix equals the product of the determinants of its blocks,
\begin{equation}
\det ( \bm S^{k k^{\prime}}) =
\det ( \bm S^{k k^{\prime}}_{\alpha })
\det ( \bm S^{k k^{\prime}}_{\beta }) \, ,
\end{equation}
and the cofactor matrix of a block-diagonal matrix has a corresponding block structure, eq.~\eqref{eq:Lowdin_NOCI} can be written as separate contributions from the $\alpha$ and $\beta$ spin channels:
\begin{align}
 \langle \Phi ^k | \hat O | \Phi ^{k^{\prime}}\rangle
 &= \det(\bm{S}^{kk^\prime} _\beta  )
 \sum _{ij \in \alpha } O _{ij} ^{k k^{\prime}} \,
 \operatorname{cof}(\bm S^{k k^{\prime}} _\alpha )_{ij} \nonumber \\
 &+ \det(\bm{S}^{kk^\prime} _\alpha  )
 \sum _{ij \in \beta } O _{ij} ^{k k^{\prime}} \,
 \operatorname{cof}(\bm S^{k k^{\prime}} _\beta )_{ij} \,.
 \label{eq:Lowdin_NOCI_unrestr}
\end{align}

Some authors~\cite{Bourne2021, Hait2021} have pointed out that the electronic contribution to the \gls{tdm} depends on the choice of origin when the ground and excited states are not orthogonal, yielding oscillator strengths that are not translationally invariant. They further point out that including the nuclear contribution to the dipole operator restores translational invariance. However, if one considers the definition of the electric dipole moment operator, eq. \ref{eq:elect_dipole}, the nuclear term is not a correction but an intrinsic part of the operator itself. Thus, as can be seen from eqs. \ref{eq:tdm} and \ref{eq:elect_dipole}, the transition dipole moment of charge neutral systems is inherently translationally invariant, even for nonorthogonal states.

Most studies where spectra are computed from \gls{oo} density functional calculations seem to use the L{\"o}wdin's formulation, which indeed provides the most straightforward route to obtain the \gls{tdm}. However, equivalent and often more efficient formulations exist, which are based on transforming the two sets of nonorthogonal orbitals to corresponding, or biorthogonal, orbitals before evaluating the matrix elements~\cite{Burton2022nonorthogonal,Burton2021,Lengsfield1981Corresponding, King1967Corresponding}. These approaches are particularly useful when many nonorthogonal coupling elements must be evaluated.

Potential issues with evaluating transition properties between nonorthogonal \gls*{oo} states were discussed by Gilbert et al. in the work presenting the \gls*{mom} algorithm~\cite{Gilbert2008} (see section \ref{sec:oo_algorithms}). They state that finite overlaps between the ground and \gls*{oo} excited states could, in principle, artificially enhance the \gls*{tdm}. However, for states of the same irreducible representation as the ground state, for which the overlap with the ground state is nonzero, the resulting oscillator strengths remained within the range predicted by configuration interaction singles (CIS) and \gls{lrtddft}. They also found that replacing Hartree–Fock with B3LYP led to very similar \glspl{tdm}, despite substantially smaller overlaps.

Bourne-Worster et al.~\cite{Bourne2021} proposed to simplify the evaluation of the length-gauge oscillator strength by first orthogonalizing the nonorthogonal \gls*{oo} states. In this approach, a set of \gls*{oo} states $\{\ket{\Phi^k}\}$ is transformed using L{\"o}wdin symmetric orthogonalization,
\begin{equation}
\ket{\tilde{\Phi}^k} =
\sum_{k^\prime}
\left(\bm{\Omega}^{-1/2}\right)_{k^\prime k}
\ket{\Phi^{k^\prime}} ,
\end{equation}
where $\Omega_{kk^\prime}=\bk{\Phi^k}{\Phi^{k^\prime}}$ is an element of the overlap matrix between the determinants. The transformed states therefore satisfy
\begin{equation}
\bk{\tilde{\Phi}^k}{\tilde{\Phi}^{k^\prime}} = \delta_{kk^\prime}.
\end{equation}
The transition dipole moments can then be evaluated between orthogonal determinants, avoiding the use of methods for nonorthogonal matrix elements. Bourne-Worster et al. applied this procedure to the HOMO--LUMO singlet transition in a set of more than 100 small neutral molecules (see also section \ref{sec:optical_spectra})\cite{Bourne2021}. They compared the symmetric-orthogonalization results with those obtained from the L{\"o}wdin formula for nonorthogonal matrix elements, including the nuclear contribution to the dipole operator as in the full molecular dipole defined above. The two approaches were found to give nearly identical oscillator strengths. However, their analysis focused on a single excitation rather than on complete spectra involving several excited states. In principle, the calculation of a spectrum requires all relevant states to be treated on a common footing. Symmetric orthogonalization can provide such a common orthonormal representation, but only if all states are included in the same orthogonalization procedure. A pairwise orthogonalization would instead generate a different orthonormal basis for each pair of states, which could complicate the comparison of transition properties across multiple excitations. It is also worth emphasizing that, from a \gls{ks} perspective, the determinants themselves are auxiliary wave functions of noninteracting reference systems. They therefore do not need to be mutually orthogonal as the exact eigenstates of the interacting electronic Hamiltonian.

Sinyavskiy \textit{et al}.~\cite{Sinyavskiy2025} have proposed an alternative route for obtaining the \gls{tdm} between a ground and an excited state from \gls*{oo} calculations without dealing with the nonorthogonality of the states. The approch is used within the SU-\gls{roks} framework, which directly computes a spin-pure open-shell singlet state and therefore avoids the need for the spin-purification formula (see section~\ref{sec:methods_roks}). Rather than evaluating transition moments directly between nonorthogonal \gls*{oo} determinants, each optimized excited state is projected onto a CIS-like expansion constructed from ground-state orbitals, and the \gls{tdm} is computed between the ground state and this CIS wave function of ground state orbitals. For an excitation from an occupied orbital $a$ to a virtual orbital $b$, the spin-adapted open-shell singlet excited state wave function can be written as (see also eq. \ref{eq:singlet_csf})
\begin{align}
    \ket{^{1}\Phi_{a}^{b,0}}
    &=
    \frac{1}{\sqrt{2}}
    \left(
    \ket{a\bar{b}} +
    \ket{b\bar{a}}
    \right), \nonumber \\ 
    &=
    \dfrac{1}{\sqrt{2}}
    \left(
    \ket{\Phi_{\bar{a}}^{\bar{b},0}}
    +
    \ket{\Phi_{a}^{b,0}}
    \right),
\end{align}
where the superscript 0 indicates that the wave function is constructed from ground-state orbitals.
The SU-ROKS excited state $k$ is then approximated as
\begin{equation} \label{eq:Luber_CI_wavef}
    \ket{\Phi^k}
    \approx
    \sum_{a,b} P_{ab}^{k}
    \ket{^{1}\Phi_{a}^{b,0}} ,
\end{equation}
where the coefficients correspond to the following projections:
\begin{equation}
    P_{ab}^k
    =
    \bk{^{1}\Phi_{a}^{b,0}}{\Phi^k}
    =
    \dfrac{
    \bk{\Phi_{\bar{a}}^{\bar{b},0}}{\Phi^k}
    +
    \bk{\Phi_{a}^{b,0}}{\Phi^k}
    }{\sqrt{2}} .
\end{equation}
In this way, the OO excited state is represented in a ground-state orbital basis. The transition moment between the ground and the excited state $k$ can then be approximated as
\begin{equation}\label{eq:Luber_f}
    \bm{\mu}^{0k}
    \approx
    \sqrt{2}
    \sum_{a,b}
    P_{ab}^k
    \expv{\psi_a^{0}}{\hat{\bm{\mu}}}{\psi_b^{0}} \, .
\end{equation}
The signs of the coefficients $P_{ab}^k$ depend on the arbitrary phases of the \gls{ks} orbitals, so a consistent phase convention between the states must be used. This projection strategy provides a practical way to avoid computing the \gls{tdm} between nonorthogonal \gls*{oo} determinants, but it also changes the representation of the excited state. This is convenient, but somewhat less natural from the perspective of \gls{oo} methods than using the \gls{oo} state directly. The approach is also limited to transition moments between the ground state and open-shell singlet excited states. 

The methods discussed above and most applications of \gls{oo} density functional approaches to calculations of absorption and emission spectra evaluate the oscillator strength in the length gauge. Recently, Shen, Fan, and Yang introduced a velocity-gauge formulation in which the momentum matrix elements are evaluated between the nonorthogonal \gls{ks} states using L{\"o}wdin's formula.~\cite{Yang2026-et} As expected for an approximate approach, the length and velocity gauges are not formally equivalent in \gls{oo} density functional calculations, because the ground and excited states correspond to separate state-specific effective Hamiltonians. For neutral molecules, Shen et al.  found velocity-gauge oscillator strengths comparable to length-gauge results.\cite{Yang2026-et} The main advantage of the velocity gauge is that the momentum operator is intrinsically origin independent, so the resulting transition moments remain origin independent also for charged systems, whereas inclusion of the nuclear contribution in the length gauge does not remove the origin dependence.

\section{Applications}\label{sec:applications}

\subsection{Rydberg excited states}
\begin{figure*}[hbt]
    \centering
    \includegraphics[width=0.85\linewidth]{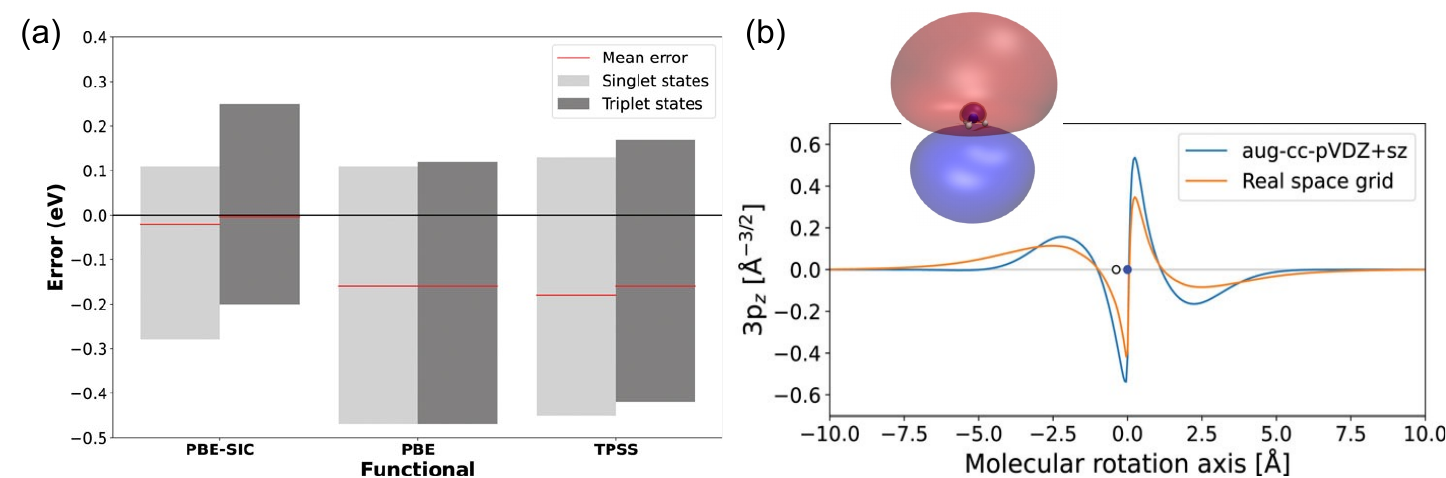}
    \caption{Performance of orbital-optimized (OO) density functional calculations for molecular Rydberg states~\citenum{Sigurdarson2023}. (a) Distribution of errors on the vertical excitation energy relative to experimental estimates for 31 singlet and triplet Rydberg excited states of water, ammonia, formaldehyde, and ethylene up to 10 eV above the ground state. The excitation energy is obtained in OO spin-unrestricted calculations with a real-space grid basis set, using the spin purification formula for the singlet states (see eq. \ref{eq:spin_purification_energy}). The error is in any case less than 0.5 eV. The GGA functional PBE and the meta-GGA functional TPSS provide similar results, systematically underestimating the excitation energy. The inclusion of explicit Perdew-Zunger self-interaction correction (SIC) in the PBE calculations, which recovers the $1/r$ long-range form of the effective potential, improves the results. (b) Ammonia $3p_z$ Rydberg orbital along the three-fold rotational axis of the molecule obtained in OO calculations with an aug-cc-pVDZ atomic basis set and a real-space grid representation, showing that the atomic basis set overly confines the orbital. Adapted with permission from A. E. Sigurdarson, Y. L. A. Schmerwitz, D. K. V. Tveiten, G. Levi, and H. J\'onsson, ``Orbital-optimized density functional calculations of molecular Rydberg excited states with real space grid representation and self-interaction correction'', J. Chem. Phys. \textbf{159}, 214109 (2023). Copyright 2023 AIP Publishing.}
    \label{fig:rydberg}
\end{figure*}

Rydberg excited states involve excitation of an electron to a highly diffuse orbital, and their energy approximately follows a Rydberg series converging toward the ionization limit. The diffuse nature of Rydberg states makes their description particularly challenging for electronic structure methods, as exemplified by a recent prediction challenge on the dynamics of the cyclobutanone molecule upon excitation to the 3s Rydberg state\cite{Janos2026}. Accurate results require a sufficiently flexible basis set representation to describe the long-range tail of the excited orbital\cite{Levi2026-ts} and their treatment in multi-configurational methods complicates the selection of the active space and the convergence of the wave function optimization. 

Rydberg states are particularly challenging for conventional \gls{lrtddft} calculations based on the adiabatic approximation. With local, semilocal, and global hybrid xc functionals, the excitation energy of Rydberg states is typically underestimated\cite{Barreiro-Lage2026, Seidu2015, Yang2011-rm, Cheng2008, Peach2008}, and Rydberg states may spuriously mix with valence or charge-transfer excitations\cite{Selenius2024}. These errors are commonly associated with the incorrect long-range behavior of the effective potential of approximate xc functionals\cite{VanMeer2014} and with the lack of orbital relaxation in TDDFT\cite{Cheng2008}. \gls{oo} density functional approaches are an attractive alternative because they naturally include orbital relaxation and their computational cost is small enough to use large basis sets.

Despite this promise, applications of \gls{oo} density functional calculations to Rydberg states have remained relatively limited. Early work focused on atoms and showed that \gls{oo} calculations with local and semilocal functionals can give a reasonable description of atomic Rydberg excitations, even in cases where \gls{lrtddft}  fails to predict bound states\cite{Yang2011, Cheng2008}. Van Voorhis and co-workers rationalized this result by noting that in an \gls{oo} excited-state calculation the effective potential is constructed from the excited-state density\cite{Cheng2008}. Although the potential obtained with a semilocal functional still eventually decays exponentially, and therefore too rapidly, the onset of this incorrect asymptotic behavior is pushed farther as increasingly diffuse Rydberg states are targeted. In contrast, conventional \gls{lrtddft}  relies on the ground-state xc potential and the too fast asymptotic decay can prevent the appearance of a proper Rydberg series\cite{Cheng2008}.

Benchmarks on Rydberg states of molecules have appeared more recently\cite{Restaino2026arxivDipole, Sigurdarson2023, Seidu2015}. Ziegler and co-workers performed \gls{oo} spin-unrestricted \gls{ks} calculations of singlet and triplet Rydberg excitations of small molecules including dinitrogen, carbon monoxide, formaldehyde, acetylene, water, and ethylene. They report relatively low \glspl{rmse} on the excitation energy with respect to experimental estimates, 0.24 eV with LDA and 0.32 eV with the BP86 functional. Interestingly, the hybrid functional B3LYP is not found to improve the results. Jónsson and co-workers later performed \gls{oo} spin-unrestricted calculations for singlet and triplet Rydberg excited states of formaldehyde, ethylene, water, and ammonia up to 10 eV above the ground state using a real-space grid basis set with a direct optimization approach (see section \ref{sec:oo_algorithms}). It is found that PBE already gives quite good results for the excitation energy, although with a systematic underestimation (\gls{rmse} and \gls{mse} with respect to experimental values of 0.24 and -0.16 eV, respectively, see Figure~\ref{fig:rydberg}). The TPSS and r2SCAN meta-GGA functionals do not systematically improve the results and show a similar tendency to underestimate the excitation energy. However, PBE with explicit Perdew-Zunger self-interaction correction\cite{Perdew1981} and complex-valued orbitals reduces the \gls{rmse} to 0.11 eV, likely due to the fact that self-interaction correction restores the correct, $-1/r$, long-range behavior of the effective potential. These results stand in contrast to those of \gls{lrtddft} calculations, which typically show a large functional dependence for molecular Rydberg states, with semilocal functionals severely underestimating the excitation energy and hybrid functionals reducing somewhat the error\cite{Barreiro-Lage2026, Seidu2015, Yang2011-rm, Cheng2008, Peach2008}. 

The study by Jónsson and co-workers also shows that the basis set representation of diffuse Rydberg orbitals is a crucial consideration\cite{Sigurdarson2023} (see Figure \ref{fig:rydberg}). Calculations with an atomic basis set including a single set of augmented diffuse functions are found to systematically overestimate the excitation energy compared to calculations with the real-space grid basis set, due to the atom-centered basis set overconfining the Rydberg orbital. The effect is most pronounced for higher and more diffuse Rydberg states, reaching deviations of about 1~eV for some states of water and ammonia.  As a result, for the most diffuse Rydberg states, \gls{oo} density functional calculations with the real-space grid basis set are found to agree better with experiment than higher-level coupled-cluster and selective \gls{ci} calculations carried out with atomic basis sets lacking sufficiently diffuse functions\cite{Levi2026-ts, Sigurdarson2023}. 

\begin{figure*}[hbt]
    \centering
    \includegraphics[width=1\linewidth]{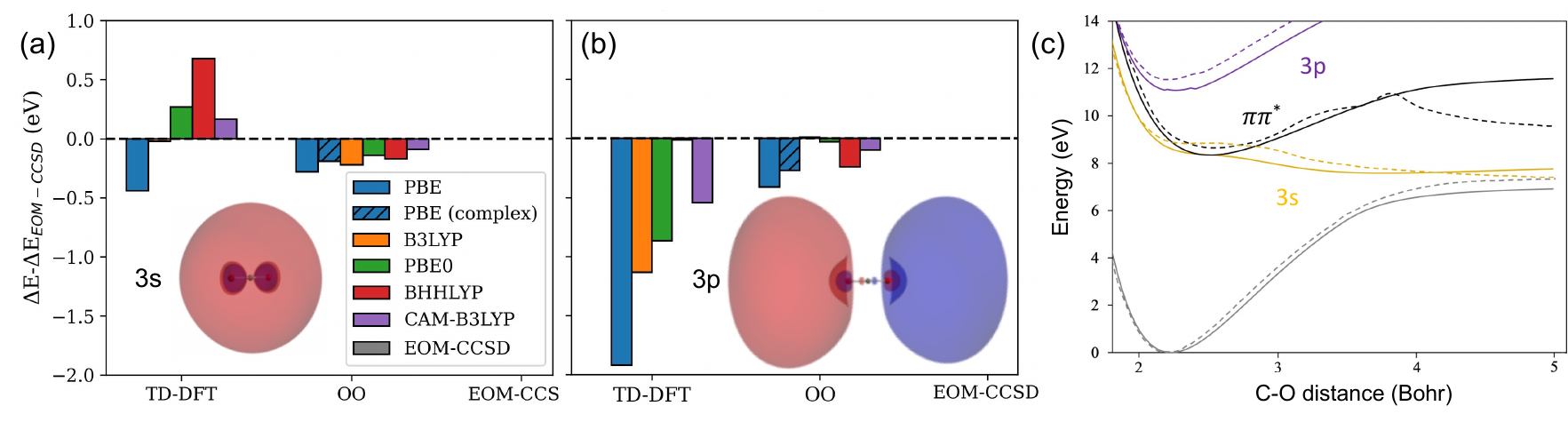}
    \caption{Performance of orbital-optimized (OO) density functional calculations for singlet valence and Rydberg excited states of carbon dioxide~\cite{Barreiro-Lage2026}. (a,b) Deviations of the vertical excitation energy from equation-of-motion coupled cluster with single and double (EOM-CCSD) reference values for a 3s (a) and 3p (b) Rydberg states obtained with TD-DFT and OO spin-unrestricted calculations using several functionals. The excitation energy from the OO calculations is spin purified. The insets show the corresponding Rydberg orbitals. The OO calculations give errors below 0.5 eV for both Rydberg states and show a weaker dependence on the functional and on the excitation than TD-DFT. (c) C–O bond dissociation energy curves for the ground state and the 3s, 3p, and $\pi^\ast$ excited states of linear CO$_2$. Solid lines show OO results with the PBE functional, while dashed lines indicate reference curves obtained using EOM-CCSD near the equilibrium geometry and multireference configuration interaction (MRCI) at long C-O distances. The OO calculations reproduce the shape of the reference curves well, including the crossing between the 3s Rydberg and $\pi^\ast$ valence states near the equilibrium ground-state geometry. Adapted from D. Barreiro-Lage, G. Levi, H. J'onsson, and T. Lamberts, ``Valence and Rydberg excited state bond dissociation curves of CO$_2$ from orbital-optimized density functional calculations’’, arXiv:2604.05802 (2026). Published under the Creative Commons Attribution 4.0 International License (CC BY 4.0).}
    \label{fig:rydberg_co2}
\end{figure*}
More recently, the electric dipole moment of molecular Rydberg states has also been investigated using \gls{oo} calculations by Restaino et al. \cite{Restaino2026arxivDipole}. This study shows that PBE also provides a reasonable description of dipole moments, with a mean absolute relative error on the magnitude of the dipole moment with respect to theoretical best estimates of $\sim$24\%, while PBE0 further improves the agreement with high-level reference values (mean absolute relative error of $\sim$15\%). In contrast to what is observed for the excitation energy, self-interaction correction with a global scaling does not improve the dipole moment and tends to overestimate its magnitude. This suggests that restoring the asymptotic tail of the potential is not sufficient and self-interaction correction likely overcorrects in regions where orbital densities overlap. Locally scaled self-interaction corrected approaches, which are currently being developed\cite{John2026arXiv, Shahi2026}, may provide a promising route to overcome these issues. The basis set sensitivity is even more pronounced for excited-state dipole moments\cite{Restaino2026arxivDipole}. A single-augmented atomic basis set can give large errors in both the magnitude and orientation of the dipole moment compared to results obtained with a plane-wave basis set, even when the excitation energy is nearly converged. Sometimes, even adding enough diffuse functions does not reproduce the plane-wave result. In such cases, the spatial extent of the density is well reproduced, but the dipole moment still differs, a discrepancy that has been attributed to the atom-centered basis lacking sufficient flexibility to describe the anisotropic redistribution of the Rydberg density\cite{Restaino2026arxivDipole}. 

More recent studies have started to address the description of potential energy surfaces of Rydberg states with \gls{oo} density functional calculations\cite{Barreiro-Lage2026, Birgisson2025dmp}. A particularly stringent test was provided by the Rydberg states of carbon dioxide along the C–O bond dissociation coordinate investigated by Barreiro-Lage et al.\cite{Barreiro-Lage2026} with OO spin-unrestricted calculations (see Figure \ref{fig:rydberg_co2}). Rydberg states in carbon dioxide include both bound and dissociative states and lie close to valence excited states, with their potential energy surfaces crossing at different molecular geometries. Again, the excitation energy is found to be affected little by the choice of functional, for both low-lying and more diffuse upper Rydberg excited states. The absolute error with respect to theoretical best estimates is below 0.5 eV for all functionals tested, including PBE, B3LYP, PBE0, BHHLYP, and CAM-B3LYP, with PBE0 providing the best results. While PBE is found to systematically underestimate the excitation energy, it reproduces the shapes of higher-level reference dissociation curves and the relative separation between excited states well, even in regions of state crossing. An important observation is that the \gls{oo} energy curves are found to have a diabatic-like character, preserving the identity of the targeted valence or Rydberg excitation along the dissociation coordinate.

\subsection{Charge transfer excited states}\label{sec:ct_states}
\begin{figure*}[hbt]
    \centering
    \includegraphics[width=0.85\linewidth]{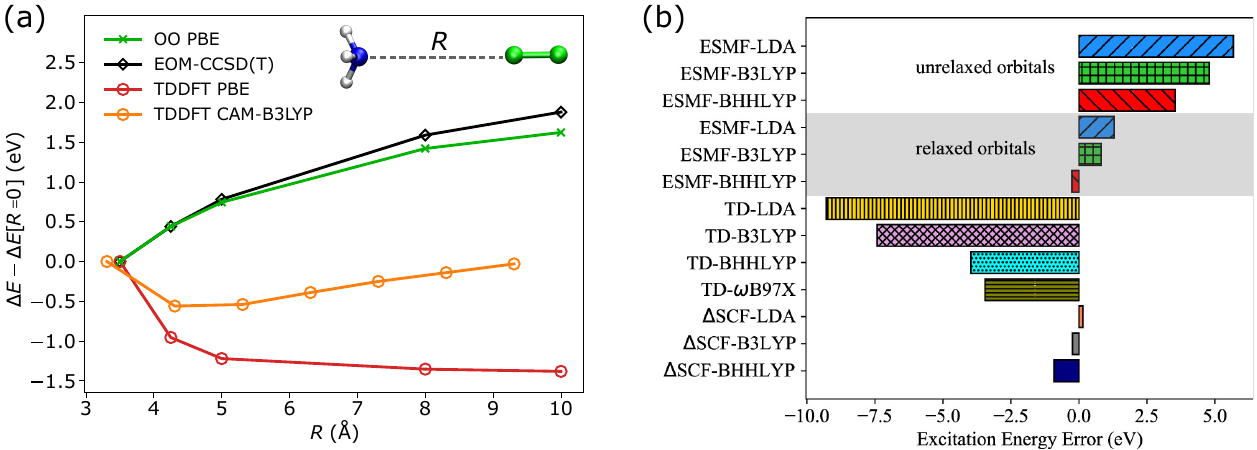}
    \caption{Comparison of orbital-optimized (\gls{oo}) and linear-response time-dependent density functional theory (LR-TDDFT) calculations of an intermolecular charge-transfer excitation in the ammonia--fluorine dimer. (a) Change in excitation energy as a function of the donor--acceptor separation. \gls{oo} PBE calculations reproduce the approximately $1/R$ dependence obtained with EOM-CCSD(T)\cite{Bogo2026,Bogo2025}, whereas LR-TDDFT gives an incorrect distance dependence with both PBE and the more advanced range-separated functional CAM-B3LYP. The \gls{oo} and LR-TDDFT calculations with PBE are from reference \citen{Schmerwitz2026}, while the LR-TDDFT CAM-B3LYP calculations are from reference \citen{Mester2022}.
    (b) Error on the excitation energy at a donor-acceptor separation of $R=6$~\AA\ with respect to EOM-CCSD for LR-TDDFT, \gls{oo} spin-unrestricted without spin purification (``$\Delta$SCF''), and excited-state mean-field restricted open-shell Kohn-Sham (ESMF-ROKS, see Table \ref{tab:roks_overview}) calculations with different functionals. For ESMF-ROKS, results obtained with and without excited-state orbital relaxation are shown. Relaxation of the orbitals significantly improves the results. Adapted with permission from L. Zhao and E. Neuscamman, ``Density Functional Extension to Excited-State Mean-Field Theory'', J. Chem. Theory Comput. \textbf{16}, 164--178 (2020). Copyright 2019, American Chemical Society.}
    \label{fig:ct_inter}
\end{figure*}
Charge-transfer excitation involves a transfer of electron density between spatially separated regions of the system. Accounting for orbital relaxation in calculations of charge-transfer excitations is therefore expected to be important. As orbital relaxation effects are missing in conventional \gls{lrtddft}, such excitations are challenging for this approach. When using local or semilocal functionals within the adiabatic approximation, in the limit of vanishing overlap between the donor and acceptor orbitals, the xc contribution to the \gls{lrtddft} coupling matrix becomes negligible, and the excitation energy essentially reduces to the corresponding KS orbital energy difference\cite{Dreuw2003,Dreuw2004}. When using approximate KS functionals, this leads to a drastically too low excitation energy due to the erroneous form of the effective potential\cite{Tozer2003}. Moreover, lack of divergence of the xc kernel of adiabatic functionals as $R\rightarrow\infty$ leads to a failure to provide the correct asymptotic $1/R$ dependence of the excited-state energy with respect to the separation $R$ between donor and acceptor\cite{Maitra2022, Hellgren2012, Dreuw2003} (see Figure \ref{fig:ct_inter}). Including exact exchange in the functional can correct these issues to some extent by adding a non-local term to the xc kernel, which can compensate for the vanishing overlap\cite{Dreuw2003,Dreuw2004}. Hence, standard range-separated hybrid functionals, such as CAM-B3LYP and $\omega$B97X, can provide accurate values of excitation energy for \textit{intra}molecular charge-transfer excitations\cite{DAntoni2025, Loos2021}. However, they might still give large errors for \textit{inter}molecular charge-transfer excitations and fail to recover the $1/R$ behavior\cite{Mester2022} unless 100\% exact exchange is added in the long range, as illustrated in Figure \ref{fig:ct_inter} for CAM-B3LYP, which has 65\% exact exchange in the long range. Optimal tuning of the range-separation parameter can improve on this\cite{Stein2009} but the approach is highly system dependent. While, in principle, frequency-dependent xc kernels beyond the adiabatic approximation can address these failures more generally, such nonadiabatic TDDFT approaches remain at an early stage of development\cite{Lacombe2023}.

\gls{oo} density functional calculations include state-specific orbital relaxation by construction, and thus are expected to avoid some of the limitations of TDDFT. Several early\cite{Briggs2015, Dreuw2003} and more recent\cite{Schmerwitz2026,Bogo2025,Bogo2024,Zhao2019,Barca2018} applications focus on intermolecular charge transfer, where the limitations of conventional \gls{lrtddft} are most severe. The excited-state charge transfer from ammonia to fluorine has become a prototype for such studies. Using the ESMF-ROKS approach described in section \ref{sec:methods_roks} with and without orbital relaxation, Zhao and Neuscamman showed that orbital relaxation is essential for getting an accurate excitation energy for this system\cite{Zhao2019}, as illustrated in Figure \ref{fig:ct_inter}. For a donor–acceptor separation of 6~\AA, relaxing the orbitals greatly reduces the error on the excitation energy, and the \gls{oo} calculations outperform \gls{lrtddft}  with all tested functionals, including the range-separated hybrid $\omega$B97X, even when the \gls{oo} calculations use the LDA. 

In an \gls{oo} calculation, the Coulomb interaction between the donor and the acceptor is included naturally through the optimized excited-state density. As a result, even a semilocal functional can recover the correct long-range physics. For the ammonia–fluorine dimer, \gls{oo} calculations with PBE reproduce the approximately $1/R$ dependence of the excitation energy on the donor–acceptor separation, in close agreement with equation-of-motion coupled cluster with single and double excitations and perturbative triples (EOM-CCSD(T))\cite{Schmerwitz2026} (see Figure~\ref{fig:ct_inter}). Similar conclusions were reached in other studies of intermolecular charge-transfer excitations. Barca et al. showed that \gls{oo} calculations recover the correct $1/R$ behavior for the ethylene–tetrafluoroethylene and bacteriochlorin–zincbacteriochlorin dimers with both the M08-HX and B3LYP global hybrid functionals, while \gls{lrtddft}  with B3LYP fails\cite{Barca2018}. Stein and co-workers compared \gls{oo} and \gls{lrtddft} calculations for molecular dimers using standard and optimally tuned range-separated hybrid functionals\cite{Bogo2025,Bogo2024}. Although range-separated hybrids and optimal tuning greatly improve the \gls{lrtddft} results, \gls{oo} calculations are found to give more accurate excitation energy values without requiring system-specific tuning. These studies indicate that for intermolecular charge-transfer excitations, \gls{oo} calculations are considerably less sensitive to the choice of functional than \gls{lrtddft}, for which the amount of exact exchange and the range-separation parameter are critical for obtaining an accurate excitation energy and the correct dependence of the energy on donor-acceptor distance.

\begin{figure*}[hbt]
    \centering
    \includegraphics[width=1\linewidth]{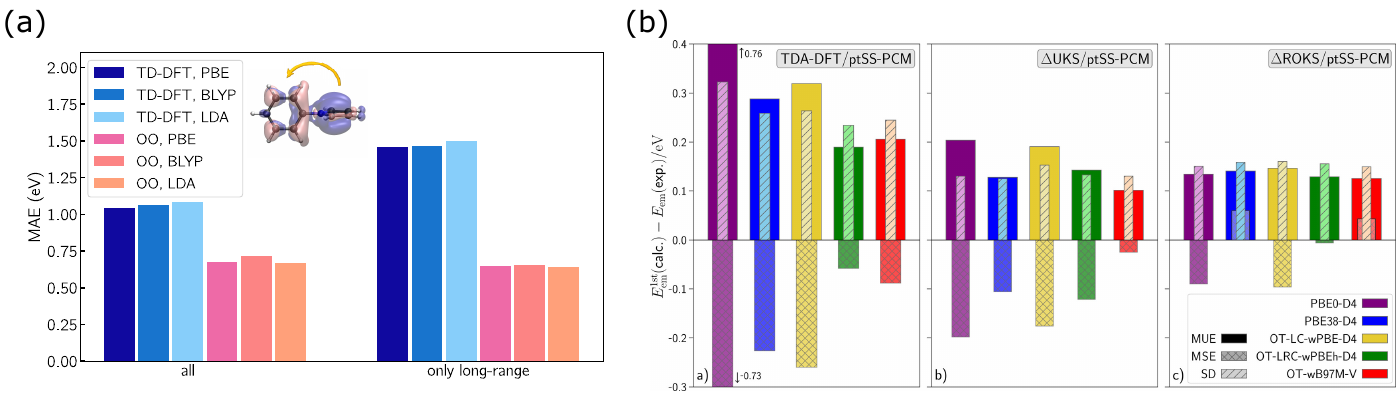}
    \caption{Performance of orbital-optimized (\gls{oo}) and linear-response time-dependent density functional theory (LR-TDDFT) calculations for intramolecular charge-transfer excitations in organic molecules. 
    (a) Mean absolute error on the vertical excitation energy with respect to CCSDT estimates\cite{Loos2021} for 27 charge-transfer excitations in 15 organic molecules. Results are shown for \gls{oo} spin-unrestricted and LR-TDFT calculations using local and semilocal functionals, both for the full set and for a subset of long-range charge-transfer excitations\cite{Selenius2024}.
    (b) Mean unsigned error (MUE), mean signed error (MSE), and standard deviation (SD) of the vertical emission energy with respect to experimental values for 27 charge-transfer states in thermally activated delayed fluorescence emitters in solution. Results are compared for LR-TDDFT with Tamm--Dancoff approximation, \gls{oo} spin-unrestricted, and spin-purification restricted open-shell Kohn--Sham (SP-ROKS, see Table \ref{tab:roks_overview}) calculations using a perturbative state-specific nonequilibrium polarizable continuum model (ptSS-PCM) for the solvent. 
    The \gls{oo} approaches are much less sensitive than LR-TDDFT to the choice of functional and to the character of the excitation. 
    Panel (a) adapted with permission from E. Selenius, A. E. Sigurdarson, Y. L. A. Schmerwitz, and G. Levi, ``Orbital-optimized versus time-dependent density functional calculations of intramolecular charge transfer excited states,'' J. Chem. Theory Comput. \textbf{20}, 3809--3822 (2024). Copyright 2024, American Chemical Society.  Panel (b) adapted with permission from T. Froitzheim, L. Kunze, S. Grimme, J. M. Herbert, and J.-M. Mewes, ``Benchmarking Charge-Transfer Excited States in TADF Emitters: $\Delta$DFT Outperforms TD-DFT for Emission Energies,'' J. Phys. Chem. A \textbf{128}, 6324--6335 (2024). Copyright 2024, American Chemical Society.}
    \label{fig:ct_intra}
\end{figure*}

For long-range intermolecular charge-transfer excitations, the improvement of \gls{oo} density functional calculations over conventional TDDFT is particularly clear, as there TDDFT tends to exhibit qualitative failures. For intramolecular charge-transfer excitations, the situation is more nuanced. In this case, the donor-acceptor separation is usually shorter, and several functionals, including range-separated hybrid functionals, have been specifically developed to improve the TDDFT description of such states. To compare the performance of \gls{oo} and TDDFT calculations for a broad range of intramolecular charge-transfer excitations, Selenius et al. performed a benchmark study of 27 excitations in 15 organic molecules, focusing mainly on local and semilocal functionals\cite{Selenius2024}. The benchmark includes excitations ranging from short-range to relatively long-range charge transfer, providing a test of whether a method can give a balanced description across different degrees of charge-transfer character. For \gls{lrtddft} with LDA, PBE, and BLYP, the error in the excitation energy is found to increase with increasing extent of charge transfer, whereas no such correlation was observed for \gls{oo} calculations with the same functionals (see Figure~\ref{fig:ct_intra}). On average, over the full set of 27 excitations, \gls{oo} calculations outperform \gls{lrtddft} for all three functionals. For the subset of 12 excitations with strong charge-transfer character, \gls{oo} calculations reduce the mean absolute error by roughly a factor of two compared to \gls{lrtddft}, as shown in Figure~\ref{fig:ct_intra}. For five representative excitations spanning the range of charge-transfer distances, the excitation energy was also calculated with B3LYP and CAM-B3LYP. With B3LYP, \gls{oo} calculations are more accurate than \gls{lrtddft} for four of the five excitations. With CAM-B3LYP, \gls{lrtddft} gave more accurate results for the three excitations with the largest charge-transfer distance, but overestimated the excitation energy for the more local excitations. Overall, the \gls{oo} calculations were found to give a more balanced description of excitations with different extent of charge transfer than \gls{lrtddft}.

Thermally activated delayed fluorescence (TADF) emitters represent an important class of systems with charge-transfer states, which are of particular interest because of their application in organic light emitting diodes\cite{Wong2017}. In a TADF emitter, triplet excitons can be thermally upconverted to emissive singlet states through reverse intersystem crossing, leading to delayed fluorescence rather than non-radiative decay from the triplet state. For this process to be thermally accessible, a small singlet–triplet gap is required. Organic molecules with charge-transfer excited states are promising TADF candidates, because the spatial separation between the donor and acceptor orbitals reduces the exchange interaction and, consequently, the energy difference between the singlet and triplet excited states. The singlet–triplet gap and the emission energy, which determines the color of the emitter, are among the key properties of a TADF candidate. Since the molecular environment can strongly affect delayed emission, for example by stabilizing polar charge-transfer states, accurate predictions require the inclusion of the effect of the environment, such as through a polarizable continuum model, for both emission energy values and singlet–triplet gaps.

Hait et al. performed a benchmark study of 27 TADF emitters in vacuum, comparing \gls{lrtddft} and SP-ROKS (see Table \ref{tab:roks_overview}) results to experimental values of singlet--triplet gap and emission energy\cite{Hait2016}. SP-ROKS was found to give more accurate results than \gls{lrtddft} with the PBE, B3LYP, and PBE0 functionals. For example, SP-ROKS gives an \gls{rmse} in the emission energy of about 0.5 eV with PBE and about 0.2 eV with either B3LYP or PBE0. The corresponding \gls{lrtddft} errors with the same functionals are approximately 1.5, 0.6, and 0.5 eV, respectively. The range-separated LC-$\omega$PBE functional gives similar accuracy for the emission energy with the two approaches, but tends to overestimate the excitation energy, indicating that the default range-separation parameter is not optimal for these systems. Overall, OO calculations with B3LYP and PBE0 give the best performance in vacuum, with PBE0 yielding the smallest errors for the singlet–triplet gaps.

Mewes and co-workers later included solvent effects in SP-ROKS as well as \gls{oo} unrestricted KS calculations of TADF emitters using a polarizable continuum model. They assessed the performance of the approaches with respect to both the singlet--triplet gap\cite{Kunze2021} and the emission energy\cite{Froitzheim2024}, using global and optimally tuned hybrid functionals and comparing the results to \gls{lrtddft} calculations including implicit solvent effects. For the singlet--triplet gaps, SP-ROKS and \gls{oo} unrestricted \gls{ks} calculations, the latter without spin purification, provide significantly more accurate results than \gls{lrtddft}, reaching chemical accuracy. SP-ROKS with the optimally tuned $\omega$B97M-V and LC-$\omega$PBE-D3 functionals gives mean absolute deviations of about 0.02 and 0.03 eV, respectively. PBE0 also performs well. SP-ROKS and \gls{oo} unrestricted calculations with PBE0-D4 give a mean absolute deviation of $\sim$0.04 eV and $\sim$0.03 eV, respectively, showing that accurate results can be obtained without optimal tuning. For the emission energy, \gls{oo} unrestricted KS and SP-ROKS calculations exhibit much weaker functional dependence than TDDFT and provide more accurate results, as shown in Figure~\ref{fig:ct_intra}. The best \gls{oo} unrestricted KS calculations give a mean unsigned error of 0.10 eV, and the spread of errors with the choice of functional is much smaller than in \gls{lrtddft}.

In multiresonance TADF emitters, instead of spatially separated donor and acceptor parts, the HOMO and LUMO orbitals involved in the charge transfer excitation are localized on alternating atoms, creating short-range charge-transfer states with a larger overlap than in traditional donor-acceptor TADF emitters \cite{Hatakeyama2016}. Mewes and co-workers \cite{Kunze2025} assessed the performance of \gls{oo} unrestricted KS and SP-ROKS calculations with respect to 35 multiresonance TADF emitters using global and range-separated hybrid functionals together with a polarizable continuum model for the solvent. The \gls{oo} unrestricted calculations are found to give more accurate singlet–triplet gaps than SP-ROKS, reaching chemical accuracy with several range-separated hybrid functionals. The authors speculate that the better performance of the unrestricted approach is connected to spin symmetry breaking, which may partly account for contributions associated with doubly excited configurations. For the emission energy, \gls{oo} unrestricted KS and SP-ROKS calculations perform similarly, with mean absolute deviations below 0.2 eV for the tested functionals, including PBE0.

\subsection{Core excitations}
As for charge-transfer and Rydberg excited states, orbital relaxation is also crucial in the description of core-level excited states\cite{Vigneshwaran2025, Sen2024, Hait2021, Besley2009}. Core excitation creates a highly localized core hole, leading to a large relaxation of the remaining electrons. Conventional \gls{lrtddft} can sometimes reproduce the qualitative features of X-ray absorption spectra, but the calculated core excitation energy is typically significantly underestimated and large empirical shifts are often required to align simulated spectra with experiment\cite{Besley2021, Besley2009, Norman2018-vg}. This error has been mainly associated with the lack of state-specific orbital optimization and with the inadequate description of exchange in the core region\cite{Vigneshwaran2025, Sen2024, Hait2021, Besley2009}. While long-range exact exchange is important for the TDDFT description of long-range charge-transfer states, core excitations instead benefit from short-range exchange exchange\cite{Song2008}. \gls{oo} density functional methods provide a natural alternative because the core-excited state is optimized variationally. Given the large errors that affect \gls{lrtddft} calculations, it is not surprising that \gls{oo} approaches have long been applied to compute core excitations and, in particular, core ionization and the associated core electron binding energy\cite{Besley2021, Norman2018-vg, Zhang2016-sb, Triguero1999-no}. Here, we focus on more recent \gls{oo} studies that assess their accuracy for the excitation energy of neutral core-level excitations and X-ray absorption spectra of molecules.

One of the early assessments of \gls{oo} methods for molecular core excitations was reported by Besley, Gilbert, and Gill\cite{Besley2009}. For first-row K-edge excitations of small molecules, \gls{oo} unrestricted B3LYP calculations with uncontracted basis functions, and without spin purification, give a mean absolute deviation of only 0.5 eV with respect to experiment, while \gls{lrtddft} with B3LYP underestimates the excitation energy by more than 10 eV on average. For second-row elements, including both 1s and 2p core excitations, the error of \gls{lrtddft} is even larger, with a mean absolute deviation of 29.4 eV, whereas \gls{oo} B3LYP reduces the mean absolute deviation to 1.5 eV. This study shows that \gls{oo} calculations can provide accurate values of core excitation energy, provided sufficient basis-set flexibility in the core region and relativistic corrections are included. Oscillator strengths for the considered transitions were also computed, but the quality of the results was not assessed quantitatively.

A more recent study by Sen and Ghosh\cite{Sen2024} compared \gls{lrtddft}, \gls{oo} density functional calculations, and multireference approaches for core excitations of open-shell light-element molecules as well as transition metal complexes. The density functional calculations used PBE0, which the authors found to perform better in \gls{oo} unrestricted calculations than the more elaborate range-separated hybrid functional CAM-B3LYP for the systems considered. For light-element K-edge excitations, the \gls{oo} PBE0 calculations give a mean unsigned error of only 0.83 eV, compared to 11.86 eV for \gls{lrtddft}. For transition metal K-edge excitations, the error of \gls{oo} density functional calculations is larger, $\sim$12 eV, but still smaller than that of restricted active space second-order perturbation theory (RASPT2), $\sim$19 eV. The authors argued that the remaining error for the transition metal systems is likely dominated by limitations of the relativistic treatment rather than by spin contamination or missing multireference effects. The study also highlighted some limitations of \gls{oo} calculations due to the use of a single determinant. Excitations involving degenerate core orbitals, as for the oxygen K-edge of nitrogen dioxide and molecular oxygen, could not be addressed in the study, and likely require a multi-determinant treatment\cite{Sen2024}.

\begin{figure}[hbt]
\centering
\includegraphics[width=\linewidth]{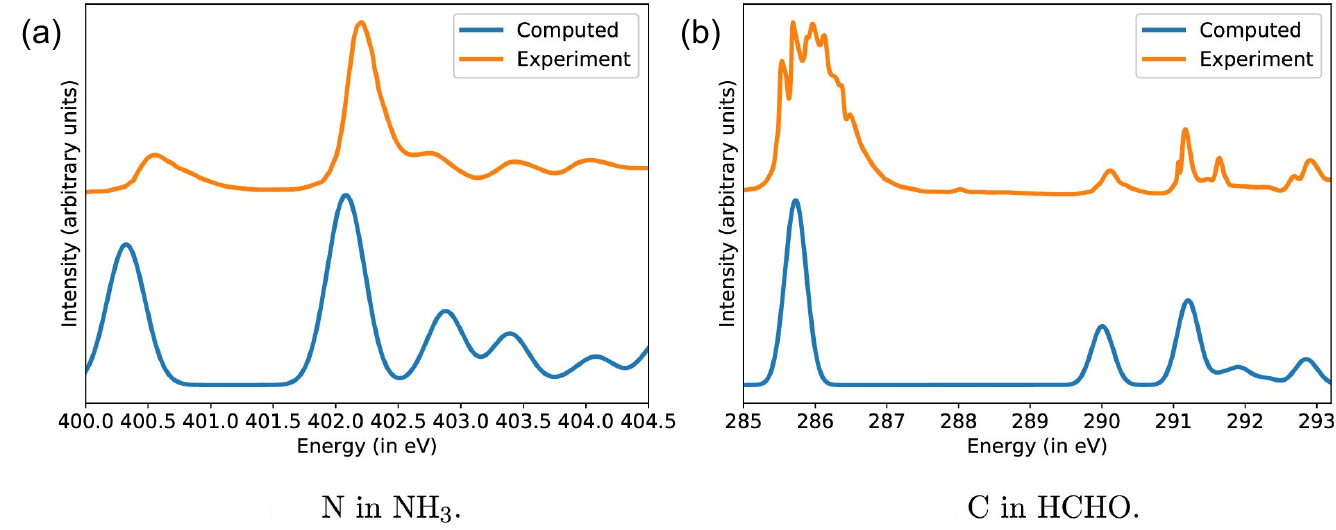}
\caption{X-ray absorption spectra computed with the spin-purification restricted open-shell Kohn-Sham (SP-ROKS) approach (see Table \ref{tab:roks_overview}) using the SCAN functional and the d-aug-cc-pCVTZ basis set, compared with experiment (without applying an empirical energy shift). The oscillator strengths for the transition intensities are computed using the KS wave function (see section \ref{sec:tdm}). (a) Nitrogen K-edge spectrum of ammonia. (b) Carbon K-edge spectrum of formaldehyde. The calculated transitions were broadened with Gaussian functions with $\sigma=0.15$ eV. The spectra reproduce the main experimental peak positions and qualitative intensity patterns.
Adapted with permission from D. Hait and M. Head-Gordon, ``Highly Accurate Prediction of Core Spectra of Molecules at Density Functional Theory Cost: Attaining Sub-electronvolt Error from a Restricted Open-Shell Kohn–Sham Approach’’, J. Phys. Chem. Lett. \textbf{11}, 775–786 (2020). Copyright 2020 American Chemical Society.}
\label{fig:Hait2020_XAS}
\end{figure}
\gls{oo} density functional methods have also been applied more recently to simulate full X-ray absorption spectra\cite{Qin2026-mf, Hait2024, Cunha2022, Hait2020core, Hait2020radicals}. For the simulations of X-ray spectra, Hait and Head-Gordon used the square gradient minimization algorithm described in section \ref{sec:oo_algorithms} in combination with SP-ROKS (see section \ref{sec:methods_roks} and Table \ref{tab:roks_overview}) to compute several core excited states and their transition intensity. The paper does not provide a detailed description of how the oscillator strengths used in the spectra were computed, only that they were obtained in a wave-function-like manner (see section \ref{sec:tdm}). For K-edge excitations of carbon, nitrogen, oxygen, and fluorine in small molecules, SP-ROKS calculations with the SCAN and $\omega$B97X-V functionals give root-mean-square errors of only 0.2--0.4 eV with respect to experiment, compared to the errors larger than 10 eV that are typical of conventional \gls{lrtddft}. The calculations were also extended to L-edge spectra of third-period elements by including spin--orbit effects perturbatively, again giving sub-eV errors with SCAN and $\omega$B97X-V. Simulated spectra for the nitrogen K-edge of ammonia and the carbon K-edge of formaldehyde reproduce the main experimental features without applying an empirical shift (see Figure~\ref{fig:Hait2020_XAS}). 

Hait et al. later extended the study to core-level spectra of open-shell radicals\cite{Hait2020radicals}. In these systems, excitation from a core orbital to a singly occupied valence orbital can be described by a single ROKS calculation, and several functionals, including SCAN, TPSS, BLYP, B3LYP, CAM-B3LYP, and $\omega$B97X-D3, give \glspl{rmse} of 0.3 eV or less for such transitions. Core excitations to empty valence orbitals are more challenging because they create three unpaired electrons and require a spin-adapted combination of several symmetry-broken determinants to describe the resulting doublet states. Hait et al. use a recoupling scheme, which optimizes several symmetry-broken determinants and combines their energy to form an approximate spin-adapted doublet excitation energy\cite{Hait2020radicals}. For the allyl radical, the SCAN spectrum reproduces the experimental C K-edge peak positions within about 0.3 eV, while TDDFT gives qualitatively incorrect positions for the higher-energy excitations to empty orbitals. The functional dependence is much smaller than in conventional \gls{lrtddft} but not negligible. SCAN and CAM-B3LYP give the best results overall, while PBE and BLYP perform less well for some higher core excitations

\begin{figure}
    \centering
    \includegraphics[width=\linewidth]{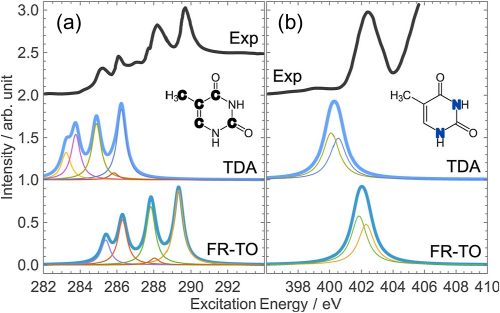}
    \caption{Carbon and nitrogen K-edge X-ray absorption spectra of thymine computed with the freeze-and-release transition-orbital (FR-TO) orbital-optimized approach (see section \ref{sec:oo_algorithms}), compared with experiment. (a) Carbon K-edge spectrum. (b) Nitrogen K-edge spectrum. The calculations were performed with the BHHLYP functional using a mixed pc-2/pcX-3 basis set. Individual transitions are broadened with a Lorentzian lineshape with full widths at half maximum of 0.6 and 1.2 eV for the carbon and nitrogen K edges, respectively. The calculated spectra reproduce the main experimental peak positions and relative spectral features more accurately than the corresponding linear-response time-dependent density functional theory calculations in the Tamm–Dancoff approximation (TDA). Adapted with permission from L. Qin and B. Suo, ``FR-TO $\Delta$SCF: A Robust and Systematic Framework for Core Excitations’’, J. Chem. Theory Comput., \textbf{22}, 4609-4625 (2026). Copyright 2026 American Chemical Society.}
    \label{fig:fr_to_spectrum}
\end{figure}
Qin and Suo recently applied a freeze-and-release \gls{oo} approach using transition orbitals from a preliminary \gls{lrtddft}  calculation (see section~\ref{sec:oo_algorithms}) to molecular core excitations\cite{Qin2026-mf}. In this approach, a preliminary \gls{lrtddft}  calculation within the TDA is used to obtain transition orbitals associated with the target excitation, which are then used to construct the initial guess for the \gls{oo} calculation. The authors addressed cases where \gls{lrtddft}  describes a core excitation as a mixture of several orbital transitions. In such cases, initializing an \gls{oo} calculation by promoting an electron from a core orbital to one selected ground-state virtual orbital, as described in section \ref{sec:initial_guess}, can become ambiguous, because different choices of the target virtual orbital may correspond to different components of the same \gls{lrtddft}  excited state. Qin and Suo showed that this situation occurs for challenging sulfur core excitations in methionine and glutathione\cite{Qin2026-mf}. At the \gls{lrtddft}  TDA level, the lowest sulfur core excitation was distributed over several virtual orbitals, so no single canonical orbital provided an obvious target for the \gls{oo} calculation. The transition-orbital construction combines these contributions into one effective excited orbital, after which the \gls{oo} calculation converges to a relaxed single-determinant solution. The approach was also used to simulate the carbon and nitrogen K-edge spectra of thymine, reproducing the main experimental peak positions and relative spectral features and outperforming \gls{lrtddft}, as shown in Figure~\ref{fig:fr_to_spectrum}. The observations by Qin and Suo raise an important open question: Are such core excited states intrinsically multi-configurational and \gls{oo} calculations manage to capture this, or does the apparent mixing mainly reflect the lack of orbital relaxation in \gls{lrtddft}  and \gls{oo} calculations uncover the single-configurational nature of the states? This point remains to be clarified by comparison with multireference methods. 

\subsection{Optical absorption and emission spectra}\label{sec:optical_spectra}
Most applications of \gls{oo} density functional calculations to optical absorption spectra of molecules have focused on assessing the oscillator strength of low-lying excited states, particularly those arising from a HOMO–LUMO transition~\cite{Yang2026-et,Toffoli2022,Vandaele2022b,Bourne2021}. In the benchmarks of Bourne Worster et al. and Shen et al. on the HOMO–LUMO excitation of 109 small closed-shell molecules, \gls*{oo} calculations are found to predict \glspl{tdm} with similar accuracy to \gls*{lrtddft} when compared with EOM--CCSD reference values, although with somewhat larger scatter~\cite{Yang2026-et,Bourne2021}. Bourne Worster et al. assessed OO calculations with CAM–B3LYP~\cite{Bourne2021}, while Shen et al. additionally tested PBE and B3LYP, finding little functional dependence in the resulting oscillator strength\cite{Yang2026-et}. Calculations by Toffoli et al.\cite{Toffoli2022} on BODIPY and aza-BODIPY dyes, where orbital relaxation effects seem to be quite important, instead show that the \gls{oo} density functional approach employed in the spin unrestricted formalism provides an improved description of the lowest HOMO-LUMO excitation relative to \gls{lrtddft}, with the reported oscillator strengths being closer to CASPT2 estimates~\cite{Toffoli2022} and PBE0 and B3LYP giving the best results. For these systems, at the \gls{lrtddft} level, the corresponding excited states often appear as mixtures of more than one orbital transition. At the \gls{oo} level however the same states seem to be effectively described by a single relaxed configuration. Overall, these works suggest that transition intensities from \gls*{oo} \gls{ks} calculations can be reliable when the target state is well represented by a single-configurational excitation. 

\begin{figure*}[hbt]
    \centering
    \includegraphics[width=0.75\linewidth]{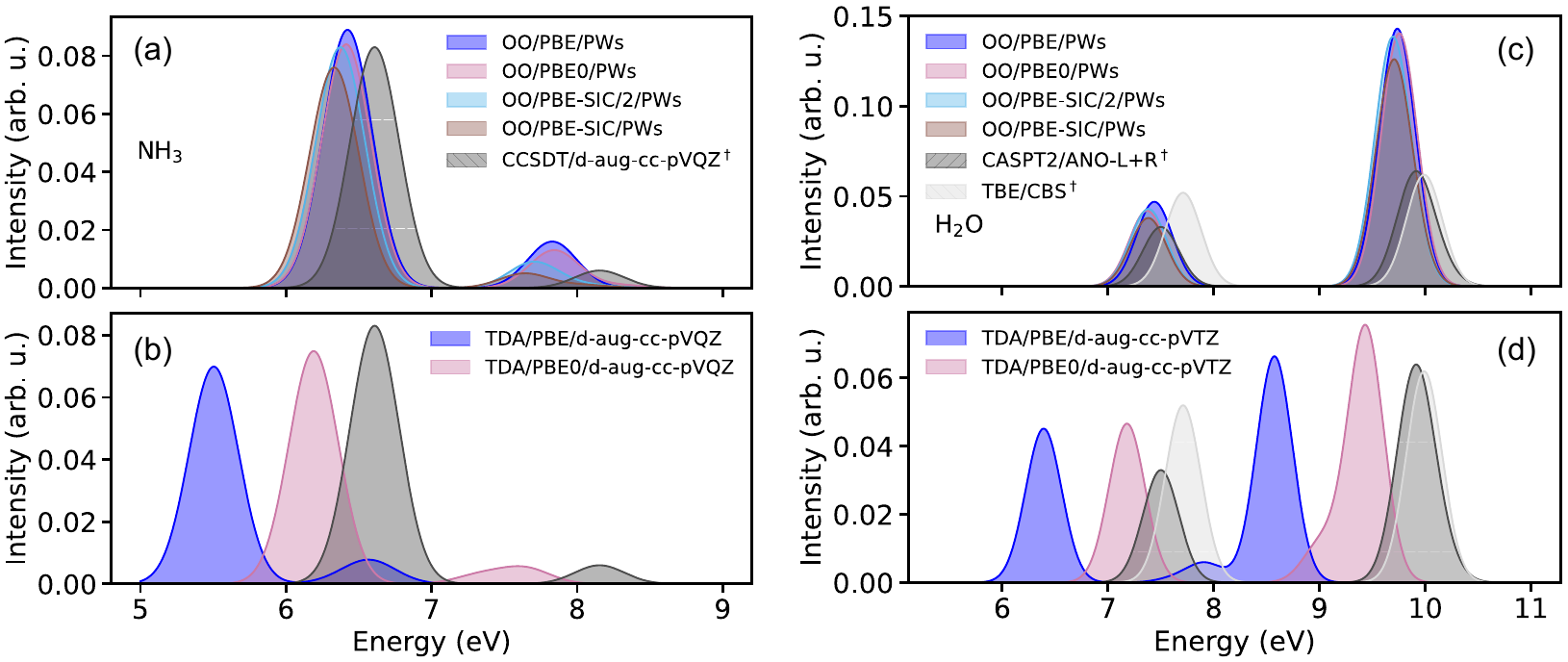}
    \caption{Optical absorption spectra from orbital-optimized (OO) density functional calculations for ammonia and water~\cite{Restaino2026arxivTDM}. (a,c) OO spectra obtained with PBE, PBE0, and PBE with explicit Perdew–Zunger self-interaction correction, using either full SIC or a globally scaled correction by a factor of 1/2. The calculations use a plane-wave (PW) basis set and are spin-unrestricted. (b,d) Corresponding linear-response time-dependent density functional theory (LR-TDDFT) spectra obtained within the Tamm–Dancoff approximation using PBE and PBE0 with a d-aug-cc-pVTZ basis set. The spectra are compared with high-level reference results, including CCSDT/d-aug-cc-pVTZ for ammonia, and MS-CASPT2/ANO-L+R\cite{Rubio2008-lk} together with theoretical best estimates extrapolated to the complete basis set limit from high-order coupled-cluster calculations (TBE/CBS)\cite{Chrayteh2021} for water. All OO calculations reproduce the peak positions well and have weak dependence on the functional, in contrast to the larger shifts and large functional dependence observed in \gls{lrtddft}. For water, the intensity of one peak is overestimated by OO because single-determinant OO calculations do not describe the mixing between nearby states of the same symmetry predicted by the higher-level calculations. Adapted from L. Restaino, D. Llorena Prieto, J. John, Y. L. A. Schmerwitz, E. \"O. J\'onsson, and G. Levi, ``Excited-state Properties Beyond the Excitation Energy from Orbital-Optimized Density Functional Calculations II: Absorption Spectra’’, arXiv:2606.13243 (2026), licensed under the Creative Commons Attribution 4.0 International License (CC BY 4.0).}
    \label{fig:spectra_water_ammonia}
\end{figure*}
Less is known about the performance of \gls{oo} density functional calculations for optical absorption spectra involving transitions beyond low-lying excitations. This limitation can be traced to the difficulty of converging higher \gls{oo} excited-state solutions, which can be high-order saddle points, as discussed in detail in section~\ref{sec:oo}. The application of \gls{oo} density functional methods to vibronic spectra also remains relatively limited compared with \gls{lrtddft}. This is notable because the vibronic structure is highly sensitive to the shape of excited-state potential energy surfaces and several studies indicate that \gls{oo} calculations can provide a robust description of such surfaces due to a balanced treatment of different electronic characters\cite{Barreiro-Lage2026, Birgisson2025dmp, Vandaele2022, Schmerwitz2022, Vandaele2022b, Malis2020, Maurer2011, Gavnholt2008}. This makes them promising for applications in vibronic spectroscopy, although work in this direction remains scarce.

A recent study by Restaino et al. has extended the assessment to optical absorption spectra involving several excited states of small molecules up to about 10 eV above the ground state, where the spectrum is dominated by Rydberg excitations~\cite{Restaino2026arxivTDM}. The calculations were spin-unrestricted, with spin purification applied to both the excitation energy and the \gls{tdm} according to eqs. \ref{eq:spin_purification_energy} and \ref{eq:spin_purification_tdm}. Several xc approximations were tested, including PBE, PBE0, and PBE with explicit Perdew–Zunger self-interaction correction. Representative spectra for ammonia and water are shown in Figure \ref{fig:spectra_water_ammonia}, together with high-level reference spectra and \gls{lrtddft} results. A trend emerging from these examples is that the \gls{oo} calculations predict the peak positions of the Rydberg spectra rather well and with weak dependence on the functional. This behavior contrasts with \gls{lrtddft}, for which the spectra are substantially more sensitive to the functional and are often red-shifted relative to the high-level references. For ammonia, the agreement between the OO and higher-level spectra extends also to the peak intensities. In this case, the relevant excited states are well described a single configuration. For the spectrum of water, however, the intensity of the peak associated with the $2\mathrm{p}_z\rightarrow 3\mathrm{s}$ excitation is significantly overestimated relative to high-level references. At the same time, the intensity of the nearby $2\mathrm{p}_x\rightarrow 3\mathrm{p}_x$ state is underestimated. This discrepancy was attributed to missing configuration mixing in the single-determinant \gls{oo} calculations. In the reference calculations, these two excitations mix as they have the same symmetry, redistributing oscillator strength between the two states. The \gls{oo} calculations do not describe this configurational mixing. This does not affect the excitation energy as the $2\mathrm{p}_z\rightarrow 3\mathrm{s}$ and $2\mathrm{p}_x\rightarrow 3\mathrm{p}_x$ excitations have similar energy, but leads to a large error on the \gls{tdm}. These results show that \gls{oo}-calculated transition intensities can be accurate for states that are single-configurational, but the approach can fail when mixing between nearby states of the same symmetry is involved.

\begin{figure}[hbt]
    \centering
    \includegraphics[width=\linewidth]{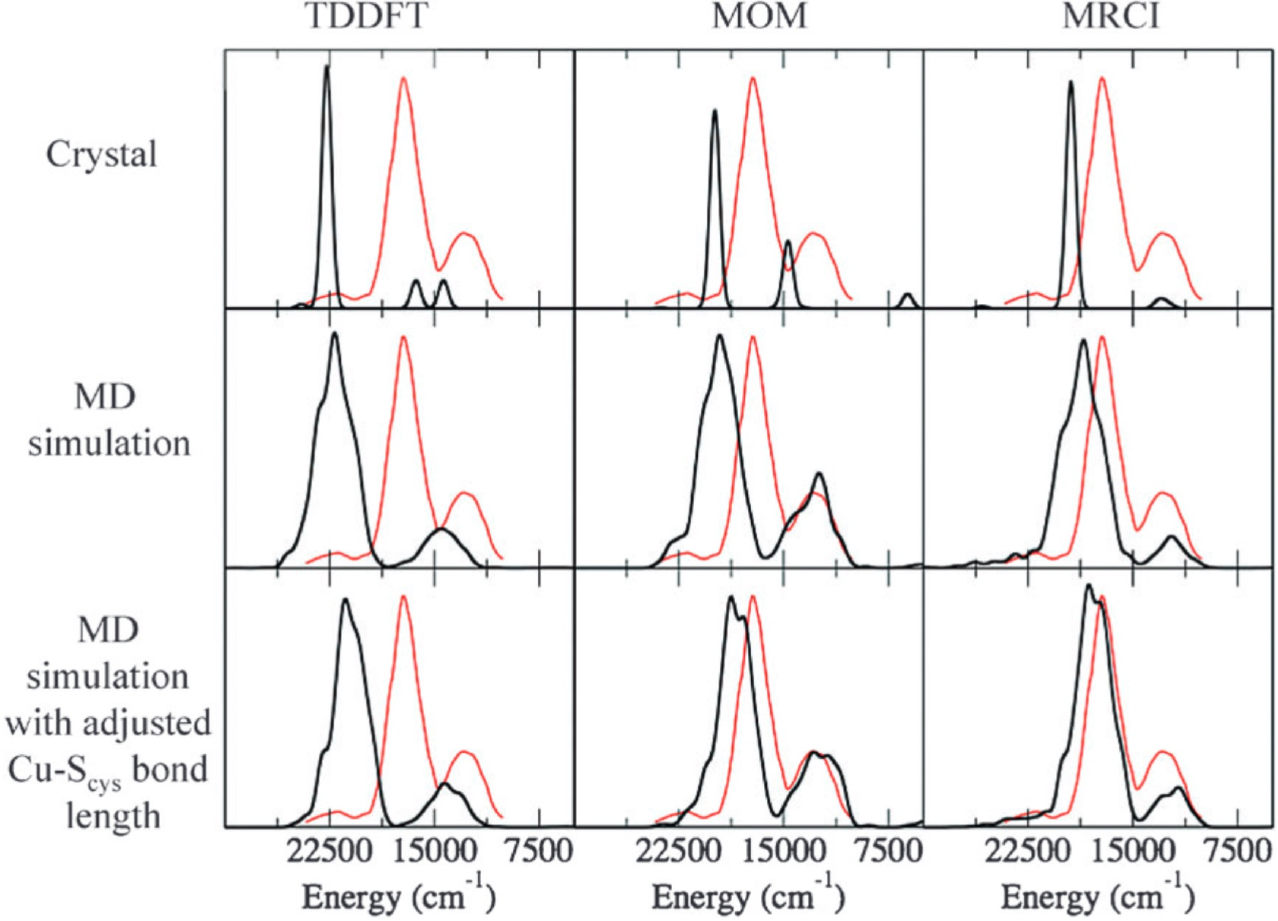}
    \caption{Computed UV/vis absorption spectrum of plastocyanin from conformational sampling~\cite{Robinson2010}. Spectra are shown for linear-response density functional theory (LR-TDDFT), orbital-optimized (OO) spin-unrestricted calculations with the maximum-overlap-method (MOM), both using a long-range-corrected hybrid functional, and multireference configuration interaction (MRCI) calculations. The three rows correspond to calculations using the crystal structure, averages over 114 structural configurations from classical molecular dynamics simulations, and averages over configurations from a modified force field in which the equilibrium Cu–S(cysteine) bond length was increased. The experimental absorption spectrum of plastocyanin~\cite{LaCroix1996-gh} is shown in red. The \gls{oo} spectrum is in better agreement with experiment compared with LR-TDDFT and reproduces the relative intensities of the bands more accurately than the MRCI spectrum. Adapted with permission from D. Robinson and N. A. Besley, Phys. Chem. Chem. Phys. \textbf{12}, 9667–9676 (2010). Copyright 2010 Royal Society of Chemistry.}
    \label{fig:Besley_spectra}
\end{figure}
An early application of \gls{oo} density functional methods to vibronic spectra is found in the work of Robinson and Besley~\cite{Robinson2010}, where \gls{oo} unrestricted KS calculations were used to compute the spectrum of the blue copper protein plastocyanin. Vibronic broadening was included through configurational sampling, by averaging vertical excitation energy values and transition intensities over structures extracted from classical molecular dynamics simulations, with each transition broadened by a Gaussian function. The authors simulated UV/vis absorption, electronic circular dichroism, and X-ray absorption spectra, including ligand-field and ligand-to-metal charge-transfer (LMCT) transitions at the copper center. For the UV/vis absorption spectrum (see Figure~\ref{fig:Besley_spectra}), the \gls{oo} KS calculations give better agreement with experiment than \gls{lrtddft} calculations performed with the same long-range corrected hybrid functional. In the \gls{lrtddft} spectrum, both the ligand-field band and the LMCT band are too high in energy. By contrast, the \gls{oo} KS spectrum places the LMCT transition closer to experiment and reproduces the relative intensities of the ligand-field and LMCT bands. The remaining discrepancy is that the LMCT band still appears slightly too high in energy, which was attributed to the sampled Cu--S(cysteine) bond length being too short~\cite{Robinson2010} (see Figure \ref{fig:Besley_spectra}). 

Another early application of \gls{oo} density functional methods to optical spectra in condensed-phase environments was presented by Briggs, Besley, and Robinson~\cite{Briggs2013}. The authors employed \gls{oo} unrestricted KS calculations of the S$_0$ and S$_1$ states of the fluorophore BODIPY, combining this treatment with molecular dynamics simulations in both the gas phase and aqueous solution. In the condensed-phase calculations, BODIPY was described with density functional calculations, while the surrounding water molecules were treated with a force field within a QM/MM framework. Absorption and emission spectra were generated from excitation energy values and oscillator strengths evaluated over ensembles of molecular dynamics configurations. Spin-purification of the energy was applied. The approach yielded ground- and excited-state structures in good agreement with CASPT2 reference results, including a nonplanar S$_1$ minimum. The simulated spectra predict a Stokes shift of approximately 0.1 eV and a blue shift of about 0.3 eV for both absorption and emission bands in water relative to the gas phase, consistent with experimental trends~\cite{Briggs2013}. The calculations also capture the broader emission band relative to absorption.

\begin{figure*}[hbt!]
    \centering
    \includegraphics[width=\linewidth]{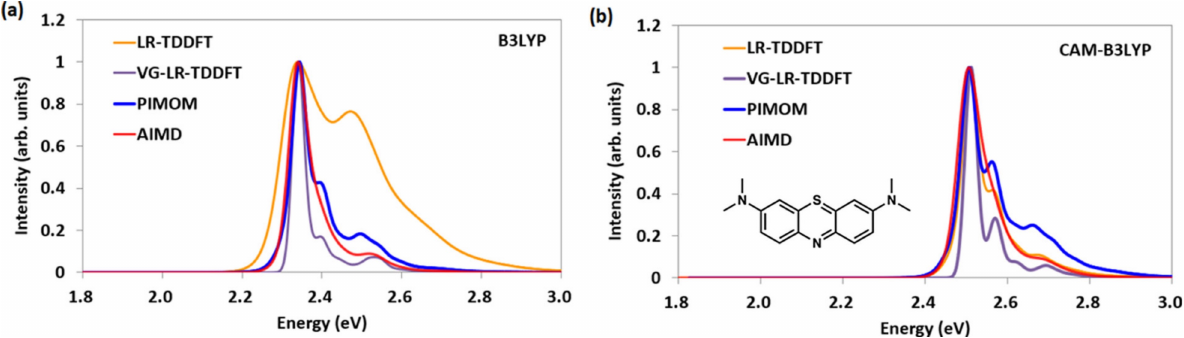}
    \caption{Vibronic absorption spectrum for the S$_0 \rightarrow$ S$_1$ transition of the methylene blue molecule computed with (a) B3LYP and (b) CAM–B3LYP. Adiabatic Hessian spectra constructed from linear-response time-dependent density functional theory (LR-TDDFT) and orbital-optimized  (OO) spin-unrestricted excited-state calculations are compared to a vertical-gradient LR-TDDFT spectrum (VG–TDDFT), and a spectrum obtained from LR-TDDFT energy-gap time-correlation functions along a ground-state ab initio molecular dynamics (AIMD) trajectory. The \gls{oo} calculations were performed using a projected initial maximum overlap method (PIMOM)\cite{Corzo2022}. With B3LYP, the adiabatic Hessian LR-TDDFT spectrum shows a pronounced vibronic shoulder that originates from mixing between the bright S$_1$ state and a nearby dark S$_2$ state near the LR-TDDFT S$_1$ minimum. This shoulder is absent in the \gls{oo} spectra, where the optimized excited state does not exhibit mixing. The difference is removed with CAM--B3LYP, which increases the S$_1$--S$_2$ gap and reduces state mixing in LR-TDDFT. Adapted with permission from A. Abou Taka, S.-Y. Lu, D. Gowland, T. J. Zuehlsdorff, H. H. Corzo, A. Pribram-Jones, L. Shi, H. P. Hratchian, and C. M. Isborn, J. Chem. Theory Comput. \textbf{18}, 3039–3051 (2022). Copyright 2022 American Chemical Society.}
    \label{fig:vibr_spectrum}
\end{figure*}

More recently, Vandaele et al.\cite{Vandaele2022} used SU-ROKS calculations (see section \ref{sec:roks} and Table \ref{tab:roks_overview}) to simulate the absorption spectrum of cyclopropanone, both in the gas phase and in aqueous solution. The spectra were generated from an ensemble of initial configurations based on a thermalized Wigner distribution for gas-phase cyclopropanone and ground-state molecular dynamics trajectories for the solvated system. For each configuration, the authors calculated the vertical excitation energy and oscillator strength of the first singlet excitation, corresponding to the HOMO--LUMO transition. The \gls{tdm} was evaluated by expanding the excited-state determinant in a basis of singly excited determinants constructed from the ground-state KS orbitals, following the procedure explained in section~\ref{sec:calculation_spectra} (see eq. \ref{eq:Luber_f}). The resulting gas-phase cyclopropanone spectrum reproduces the position of the experimental absorption maximum well, with maxima red-shifted by only 7 and 15 nm in the length and velocity gauges, respectively. In aqueous solution, the spectrum appears blue-shifted relative to the gas phase, which the authors attributed to hydrogen bonding and electrostatic stabilization of the ground state relative to the n--$\pi^\ast$ excited state.

A more explicit treatment of vibronic structure with an \gls{oo} density functional method was reported by Abou Taka et al.\cite{Taka2022}. The authors used \gls{oo} unrestricted calculations of the lowest excited state of the methylene blue molecule and compared the resulting S$_0 \rightarrow$ S$_1$ vibronic absorption spectrum with spectra obtained from \gls{lrtddft}. Figure~\ref{fig:vibr_spectrum} compares four ways of constructing the spectrum: adiabatic Hessian spectra based on either \gls{lrtddft}  or \gls{oo} calculations, a vertical-gradient \gls{lrtddft}  spectrum, and a spectrum obtained from \gls{lrtddft} energy gap time-correlation functions along a ground-state ab-initio molecular dynamics (AIMD) trajectory. With B3LYP, the adiabatic Hessian \gls{lrtddft}  spectrum shows a pronounced vibronic shoulder that is absent from the other three spectra. The authors traced this feature to mixing between the bright S$_1$ state and a nearby dark S$_2$ state near the \gls{lrtddft}  S$_1$ minimum, which changes the character of the adiabatic excited-state surface used to compute the Hessian. By contrast, the \gls{oo} spectrum remains close to the vertical-gradient and AIMD-based spectra, indicating that the \gls{oo} state has a diabatic-like electronic character. This behavior is strongly functional dependent for \gls{lrtddft}. When CAM--B3LYP is used, a larger S$_1$--S$_2$ gap reduces state mixing and the adiabatic Hessian, vertical-gradient, and AIMD-based spectra become much more similar. The \gls{oo} calculations are less affected by this change of functional, as the optimized S$_1$ state retains the electronic character with both B3LYP and CAM--B3LYP.

\section{Concluding remarks and perspectives}
Variational orbital-optimized density functional calculations provide a conceptually simple, time-independent route to electronic excited states. In these methods, excited states are computed as stationary points higher in energy than the ground state on the electronic energy surface defined by a density functional approximation, making them a natural extension of ordinary ground-state DFT calculations. Although fully variational \gls{oo} density functional methods have been used for a long time, they have been explored much less extensively than TDDFT. In part, this is because a general Hohenberg--Kohn theorem for individual excited states is lacking\cite{Gaudoin2004}. As discussed throughout this review and emphasized elsewhere\cite{Dupuy2026, Trushin2026Chemrxiv}, however, \gls{oo} approaches rest on solid theoretical foundations, with connections to several formal density functional theories\cite{Fromager2025, Kowalczyk2011, Ayers2012,Gorling1999}. Further developing these connections could provide a systematic path for improving \gls{oo} methods beyond their current capabilities. In particular, from the ensemble DFT perspectve, it has recently been shown that individual excited states can be obtained from a stationarity principle with respect to the ensemble density\cite{Dupuy2026,Fromager2025,Gould2025pra}. Exploring the connection with the exact ensemble framework from an analytical and numerical point of view can be a promising route.

A second factor that has limited the broader use of \gls{oo} density functional methods until so far is a practical one. As illustrated here, excited-state \gls{oo} solutions are typically saddle points on the electronic energy surface, so they are more difficult to locate than the ground-state minimum. As a result, excited-state \gls{oo} calculations cannot rely blindly on standard ground-state SCF algorithms. They require procedures that prevent variational collapse,  and distinguish the target excited-state solution from other nearby stationary points. Recent developments in algorithms designed specifically to locate saddle points on the electronic energy surface\cite{Qin2026-mf, Schmerwitz2026,Bogo2025,Carter2020,Hait2020} have significantly improved the robustness of these calculations and are moving \gls{oo} methods closer to routine use. Further progress will require a more detailed understanding of the electronic energy landscape generated by density functional approximations, which remains less well-understood than for many wave function methods\cite{Marie2023,Burton2022}. To make \gls{oo} methods competitive with TDDFT, where multiple states are computed simultaneously, the calculations would especially benefit from more robust and automatized choices of the initial guess together with strategies for global exploration of the electronic energy surface\cite{Dong2020,Thom2008}, rather than attempting to converge specific excited states one at a time, as is now commonly done. 

Beyond these practical optimization challenges, the predictive accuracy of \gls{oo} density functional methods also needs to be improved. As shown in this review, state-specific orbital relaxation often gives significant advantages over TDDFT, providing a more balanced description across states of different electronic character. However, in many applications, the remaining errors are still above chemical accuracy. At present, \gls{oo} methods are especially attractive for large systems, and screening applications where trends may be more important than chemical accuracy. To become more broadly competitive as predictive excited-state methods, especially as multireference approaches continue to expand toward larger systems thanks to advancements in technologies, algorithms, and machine-learning assisted approaches\cite{giuliani2026arXiv, Levi2026-ts, Shayit2025}, \gls{oo} density functional methods will require methodological improvements aimed at accuracy rather than only convergence.

One important direction is to move beyond the single-determinant KS description used in most current \gls{oo} calculations. Several restricted open-shell KS formulations have been developed for open-shell singlet excited states, which have multi-determinant configuration state function, and some have been extensively benchmarked, but it remains unclear which formulation is most reliable across different classes of excitations. In our view, a central difficulty is that most density functional approximations are not designed to describe the exchange-correlation energy associated with open-shell singlet CSFs. More broadly, a practical \gls{oo} density functional approach capable of treating arbitrary multi-determinant CSFs is still missing. This is despite the fact that the formal KS theory itself can be formulated for general CSFs\cite{LoosGiarrusso2025,Gunnarsson1976}. Promising steps in this direction have been reported\cite{Trushin2026Chemrxiv, Gould2026}, but these developments remain at an early stage and may need to be supported by functionals designed explicitly for multi-determinant excited states.

Improving the predictive accuracy of \gls{oo} density functional methods will also require renewed attention to functional development. Most excited-state functional development has been driven by the needs of TDDFT, where different types of excitations often require different types of functionals. For example, long-range corrected functionals such as CAM-B3LYP\cite{Yanai2004} improve TDDFT-calculated intramolecular charge-transfer excitations, while optimally tuned range-separated functionals are often needed for long-range charge transfer. A different class of excitations, core excited states, instead requires short-range exact exchange within TDDFT\cite{Song2008}. The applications reviewed here show that \gls{oo} methods exhibit a markedly weaker dependence on the functional choice and on the type of excitation, because state-specific orbital relaxation removes the imbalance that TDDFT must compensate through the response kernel and the xc potential. A relatively simple global hybrid such as PBE0\cite{Adamo1999-et}, for example, gives good results for several of the Rydberg, charge-transfer, valence, and core-excitations discussed here, with typical errors of a few tenths of an eV\cite{Restaino2026arxivTDM, Restaino2026arxivDipole, Barreiro-Lage2026, Kunze2025, Sen2024, Hait2020core}. In many cases, more elaborate range-separated and optimally tuned functionals do not lead to a systematic improvement. Future functional development to further improve on the predictive power of \gls{oo} density functional methods should take into account not only ground-state properties, but also excited-state properties that can be obtained from time-independent \gls{oo} calculations. For example, the excitation energy can be affected by an imbalance of self-interaction errors between the ground and excited states\cite{Schmerwitz2022}. This motivates the development of intrinsically self-interaction free functionals for \gls{oo} excited-state calculations. A complementary, more practical, promising direction is the development of locally scaled self-interaction corrections\cite{John2026arXiv, Shahi2026}.

Another fundamental limitation is the lack of established multi-configurational \gls{oo} density functional methods. Most current approaches are built around the optimization of a single excited-state configuration, or a spin-adapted CSF associated with one dominant orbital excitation, as in the case of the ROKS methods. This can be effective when the target state has clear single-configurational character. However, excited states can be intrinsically multi-configurational, e.g., in the vicinity of conical intersections. Single-configurational \gls{oo} methods may still give reasonable properties, such as the excitation energy, in such complicated cases because the interacting states are close in energy. However, other properties, such as dipole and transition dipole moments, nonadiabatic couplings, and the topology of excited-state energy surfaces near conical intersections can depend more sensitively on the mixing between configurations. In these cases, a single optimized configuration may not be sufficient, as illustrated by the optical absorption spectrum of water in Figure \ref{fig:spectra_water_ammonia}, where OO calculations significantly overestimate the intensity of the second lowest-lying peak due to lack of mixing between different configurations. There have been recent attempts to incorporate multi-configurational character through fractional orbital occupations\cite{Sinyavskiy2025}, which provide a promising route, but their accuracy and range of applicability remain to be established. In such schemes, the fractional occupations, which determine a mixing of excitations, are inferred from \gls{lrtddft} calculations. \Gls{lrtddft} can, however, give qualitatively incorrect state mixing due to a lack of orbital relaxation\cite{Selenius2024, Taka2022}, so using TDDFT to define the multi-configurational character may reintroduce the same imbalance that \gls{oo} methods are designed to avoid. For excited-state \gls{oo} density functional methods to become broadly competitive tools, they will need to evolve beyond the optimization of single configurations toward genuinely variational multi-configurational approaches.

\begin{acknowledgments}
The authors thank Elvar Ö. Jónsson, Emmanuel Fromager, Hannes Jónsson, and Yorick L. A. Schmerwitz for insightful and stimulating discussions. E.S. and G.L. acknowledge support by the Icelandic Research Fund (grant nos. 239678 and 2511544). G.L., G.G., and L.R. acknowledge support from the ERC under the European Union's Horizon Europe research and innovation programme (grant no. 101166044, project NEXUS). Views and opinions expressed are however those of the author(s) only and do not necessarily reflect those of the European Union or ERC Executive Agency. Neither the European Union nor the granting authority can be held responsible for them. 
\end{acknowledgments}

\section*{Data Availability Statement}
The data that support the findings of this study are available from the corresponding author upon reasonable request.

\appendix

\bibliography{main}

\end{document}